\documentclass[
superscriptaddress,
reprint,
amsmath,
amssymb,
aps,
physrev,
]{revtex4-2}
\usepackage{booktabs}
\usepackage{threeparttable}
\usepackage{graphicx}
\usepackage{dcolumn}
\usepackage{xcolor}
\usepackage{bm}

\begin{document}

\title{\textbf{A Learned Atomic Bonding Scale beyond Electronegativity}}

\author{Jose M. Munoz}
\email[Corresponding author: ]{josemm@mit.edu}
\affiliation{Massachusetts Institute of Technology, Cambridge, Massachusetts 02139, USA}

\author{Mateo Londoño}
\email[Corresponding author: ]{mateo.londono@stonybrook.edu}
\affiliation{Department of Physics and Astronomy, Stony Brook University, Stony Brook, New York 11794, USA}

\author{Ronald F. Garcia Ruiz}
\affiliation{Massachusetts Institute of Technology, Cambridge, Massachusetts 02139, USA}

\author{Jesús Pérez-Ríos}
\affiliation{Department of Physics and Astronomy, Stony Brook University, Stony Brook, New York 11794, USA}

\date{\today}

\begin{abstract}

Electronegativity organizes chemical bonding, but it does not fully capture how atomic electronic structure shapes molecular properties. It cannot distinguish elements that have similar electronegativity values but different valence shells. Here, we show that a second variable can account for these differences. A symmetry-preserving embedding model trained on the constants of 262 diatomic molecules recovers electronegativity as its leading direction. The remaining contribution is built from one term for each atom. It tracks the number of electrons outside a closed shell and places coinage metals beside alkali metals and halogens. The representation retains bond-order information that electronegativity alone does not capture, and includes calibrated uncertainties that identify where the periodic table remains unmeasured. To test generalization, we used it to rank 4,851 diatomics for laser cooling. Among the candidates, the molecule AgAs was predicted to have a Franck--Condon factor of 0.997.
The proposed scale therefore offers a chemically interpretable way to guide expensive quantum chemistry calculations in the search for molecules with desired properties, including molecules suitable for laser cooling.

\end{abstract}

\maketitle

\section*{Introduction}

The relationship between the properties of a molecule and those of its
constituent atoms is one of the central problems of chemical physics.
This is the case even for diatomic molecules, where constants such as
the equilibrium bond length, dissociation energy, and vibrational
frequency are not described by a known closed form in terms of atomic
properties. Relating them to the atoms requires an understanding of how
the atoms bond.

For most of a century this understanding has rested on one atomic
scale~\cite{Pauling1932,Coulson1952}. Electronegativity was itself extracted from
molecular data, and the Van Arkel--Ketelaar (VAK) triangle organizes
the metallic, covalent, and ionic limits of bonding through two
combinations of it. For a diatomic molecule AB, with $\chi_A$ and $\chi_B$
the Pauling electronegativities of atoms A and B, these are the mean $\bar{\chi}=(\chi_A+\chi_B)/2$ and the
absolute difference
$\Delta\chi=|\chi_A-\chi_B|$~\cite{AllenCapitani1993,MeekGarner2005}.
We use the Pauling scale throughout.
This picture gives chemical intuition for the bond. Molecules near the top vertex of the triangle show an ionic bond, and those near the right vertex a covalent bond. Those near the left vertex are metallic, which for
diatomics often corresponds to weak, van der Waals-like binding. The
triangle nonetheless remains qualitative. Because it depends only on
electronegativity, it cannot distinguish atoms that have similar
electronegativities but different valence occupancy or shell structure and other properties. As a result, the triangle, and with it the general chemical intuition, fails in some scenarios to determine fundamental electrostatic properties of diatomics, such as the dipole moment.

Statistical learning approaches have become increasingly useful for a
variety of open problems in
chemistry~\cite{Butler2018,Fedik2022,Raghunathan2022,madika2025artificial,passaro2025,wedig2025remdi}. Representations learned from such data can recover the structure of the
periodic table~\cite{Zhou2018,Tshitoyan2019,Willatt2018}, and in nuclei
they recover shell structure and the magic
numbers~\cite{Kitouni2023,Kitouni2024}.
The dipole moment of a diatomic molecule can be predicted from atomic
properties, accurately enough to cover the periodic table
despite the scarcity of available
data~\cite{LiuPerezRios2020,LiuPerezRios2021,Ibrahim2024}. These models also show
that $\Delta\chi$ is not enough to determine molecular polarity, and
that the relation between the VAK coordinates and molecular properties
is nonlinear~\cite{Elhalawani2026}. However, these approaches predict
one property at a time. They do not say which atomic information the
bond selects, or how to read that information back in chemical terms.

Here we construct a representation of diatomic bonding from atomic
properties. The representation is distilled
from Gaussian-process estimates of molecular dipoles for $4{,}851$
diatomics and fine-tune it on $262$ reference molecules with measured
or calculated dipole moments and spectroscopic constants. The dominant
direction of the representation recovers the electronegativity
ordering following the VAK triangle plus a residual that defines a coordinate $\xi$ for each molecule. This new coordinate depends on the atomic contribution from each atom, $e(X)$, introducing a complementary learned atomic scale beyond electronegativity for every atom. It is defined for $99$ elements and follows valence occupancy. It is raised by a single electron outside a closed shell or a single vacancy in one, and it is lowered by filled $s$ and $d$
subshells and by open $f$ shells. Within this new scale, hydrogen, the light alkali metals, and the coinage metals, which share a single valence $s$ electron, take the largest values. 

%close to the light alkali metals and to the halogens, reflecting their
%analogous valence-shell configurations with a single electron or
%vacancy relative to a closed shell. Shell structure enters the model as
%input, so the content of $e(X)$ is which of those inputs the bonding
%constants select, and with what weight.

%The representation carries information that electronegativity does not. Formal
%bond order is predicted from the representation where the VAK
%coordinates alone cannot predict it. The decoder departs from the
%electronegativity prior only for hydrides and for halides and
%chalcogenides of heavy metals, which are the cases in which
%$\Delta\chi$ is known to be a poor guide to bonding. The predictions
%come with calibrated uncertainties that hold on a fixed held-out set,
%and the molecules that depart most from the learned trends include many
%hydrides. These molecules point to where new measurements or
%calculations are most useful.

The search for molecules suitable for laser cooling provides a test of
whether the representation carries information about a property it never saw.
Since the laser cooling and magneto-optical trapping of
SrF~\cite{Shuman2010,Barry2014}, these techniques have been extended to
species including CaF, YO, CaH, BaF, YbF, and
AlF~\cite{Anderegg2017,PadillaCastillo2025}, together with a growing set
of polyatomic molecules containing optical cycling
centers~\cite{Augenbraun2023}. These molecules serve as platforms for
quantum computation in optical-tweezer arrays~\cite{Holland2023},
ultracold chemistry~\cite{Liu:2020}, and precision tests of fundamental
symmetries~\cite{jadbabaie2026radioactive}. The set of suitable species
remains small because favorable vibrational branching, a short
excited-state lifetime, and weak losses through other electronic states
must hold at once~\cite{DiRosa2004,FitchTarbutt2021}.

%In the diatomic
%molecules laser cooled to date, the optical transition involves, in some cases, a
%valence electron that is localized mainly on one atom and contributes
%little to the bond~\cite{DiRosa2004,IsaevBerger2016,FitchTarbutt2021}. The ground- and
%excited-state potentials then have similar equilibrium bond lengths,
%giving near-diagonal Franck--Condon factors. This condition is set by
%the ground-state bond, which the representation describes. Confirming
%it requires explicit excited-state calculations, which are too costly
%to apply to all combinations of atoms across the periodic table.

We therefore rank 4,851 neutral diatomics using a classifier based exclusively on atomic descriptors and ground-state molecular information. We subsequently assess selected candidates through high-level ab initio calculations of their ground- and excited-state properties, providing an independent test of the ability of the learned representation to identify potentially favorable optical-cycling transitions. This result demonstrates that a representation learned without excited-state information can guide the identification of promising optical-cycling candidates, while calculations for other selected molecules reveal the limitations of ground-state screening. Moreover, with other training targets, the same approach could prioritize molecules for other applications.

%We therefore rank the $4{,}851$ neutral diatomics by similarity of their ground-state bonding to identify optical-cycling species, using a classifier that sees only the ground-state representation. As a result, we identify new candidates for laser cooling and further test our predictions against high-level ab initio electronic structure calculations of the ground and excited states, thereby confirming the validity of the classification. We find that AgAs has a diagonal $X\,^3\Sigma^-\rightarrow A\,^3\Pi$ transition with $f_{00}=0.997$.  Furthermore, we can characterize the limitations of the predictions using molecules like AgTe and AuW, which illustrate two ways the ground-state picture fails: spin--orbit structure and a delocalized active electron.

% This work is organized as follows. We construct the representation and
% the coordinate $\xi$ in Sec.~\ref{sec:plane}, and test its prediction
% of molecular constants and their uncertainties in
% Sec.~\ref{sec:generalization}, where we also show where the reference
% data are missing. We define the atomic scale $e(X)$, compare it with
% existing scales, and examine what it encodes in Sec.~\ref{sec:encodes}.
% We then use it to select optical-cycling candidates and analyze their
% potential-energy curves, orbitals, and Franck--Condon factors against
% controls in Sec.~\ref{sec:discovery}, before discussing the scope of
% the results in Sec.~\ref{sec:discussion}.

%%%%%%%%%%%%%%%%%%%%%%%%%%%%%%%%%%%%%%%%%%%%%%%%%%%%%%%%%%%%

\section*{Results}

\subsection*{A symmetry-constrained representation of diatomic bonding}
\label{sec:plane}

We construct a representation of diatomic bonding by learning the
relation between atomic properties and four molecular properties:
the dipole moment $|\mu|$, dissociation energy $D_e$, equilibrium
bond length $R_e$, and equilibrium vibrational harmonic frequency $\omega_e$. The dipole moment measures charge separation, the dissociation
energy gives the depth of the potential well or the strength of the interaction, and the equilibrium
bond length and equilibrium vibrational harmonic frequency constrain its position and
curvature. Learning them together allows us to examine which atomic
information describes these different aspects of bonding within a
common representation.

The model is summarized in Figure~\ref{fig:arch-simple}. The full
architecture and its two stages of training are shown in
Supplementary Fig.~\ref{fig:arch}. The inputs consist of atomic quantities, including
electronegativity, ionization potential, electron affinity, group,
period, and valence-electron counts. No molecular geometry is supplied as an input, and no excited-state molecular quantities are used to train the
representation. Additional details are provided in the Methods and
Supplementary Information.

\begin{figure}[t]
  \centering
  \includegraphics[width=\linewidth]{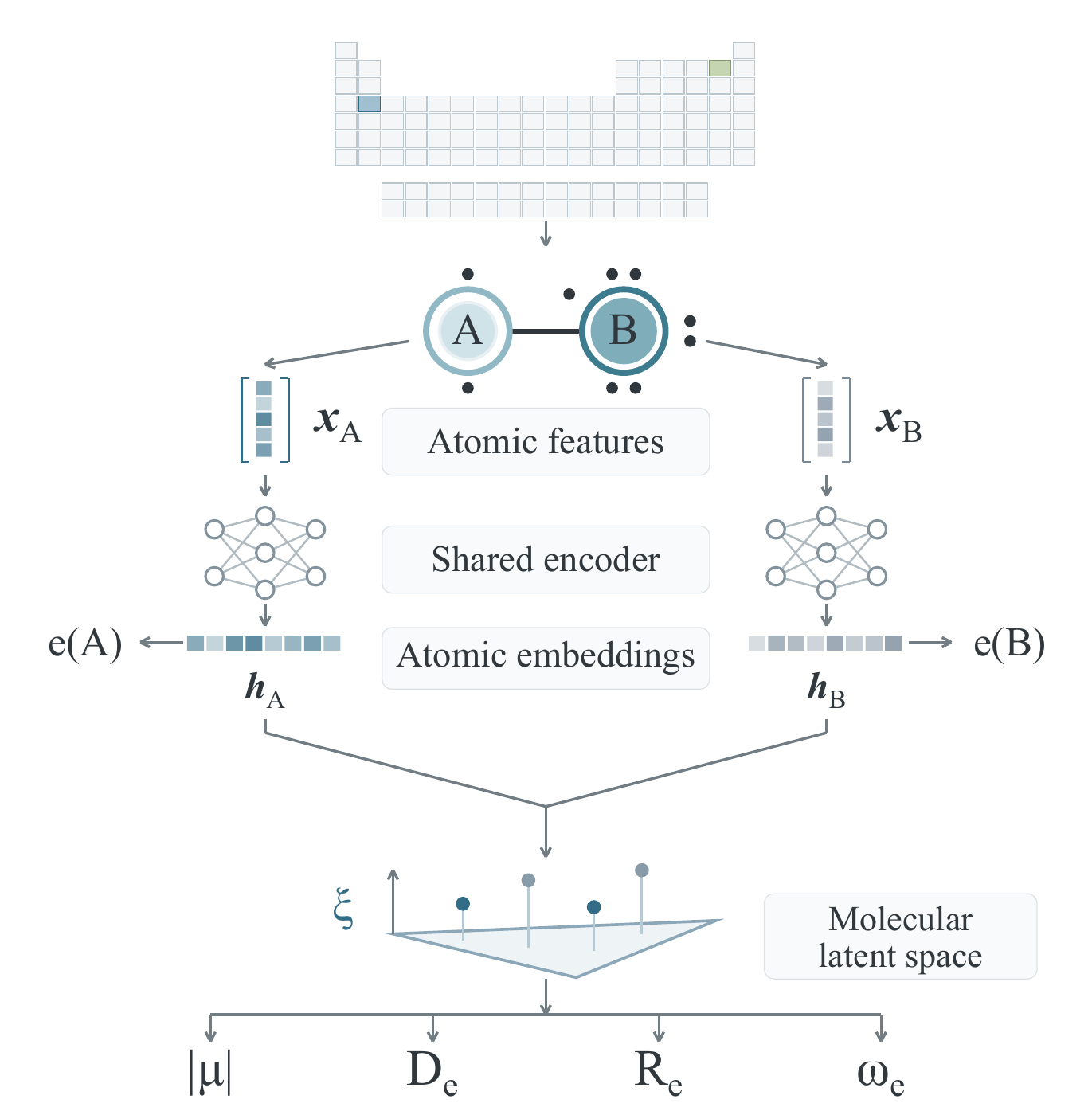}
  \caption{\label{fig:arch-simple}
  Learning a representation of diatomic bonding from atomic
  properties. A shared encoder maps the features of each atom,
  $\mathbf{x}_A$ and $\mathbf{x}_B$, to an atomic representation,
  $\mathbf{h}_A$ and $\mathbf{h}_B$, which gives the atomic
  contributions $e(A)$ and $e(B)$. The two atomic representations
  form a molecular representation that extends the VAK triangle
  by the residual coordinate $\xi$ and predicts $|\mu|$, $D_e$,
  $R_e$, and $\omega_e$. The full architecture is shown in
  Supplementary Fig.~\ref{fig:arch}.}
\end{figure}

Two considerations guide the model construction. First, the
predicted molecular properties must be invariant under exchange
of the two atoms~\cite{Zaheer2017}. We apply the same encoder to both atoms (AB), and denote their posterior-mean representations by $\mathbf{h}_A$
and $\mathbf{h}_B$ for atoms A and B, respectively. These representations are combined into symmetric and
antisymmetric channels,
\begin{equation}
 \mathbf{z}_{\mathrm{sym}}
 =\tfrac{1}{2}(\mathbf{h}_A+\mathbf{h}_B),
 \qquad
 \mathbf{z}_{\mathrm{diff}}
 =\mathbf{h}_A-\mathbf{h}_B.
 \label{eq:invariant-latent}
\end{equation}
Each channel has $16$ dimensions. Exchanging the atoms leaves
$\mathbf{z}_{\mathrm{sym}}$ unchanged and reverses the sign of
$\mathbf{z}_{\mathrm{diff}}$. We evaluate the decoder for both
signs of the antisymmetric channel and average its output, so the
molecular predictions satisfy the exchange symmetry exactly.
The two channels retain information about both the mean atomic
properties and the differences between the constituent atoms.

Second, we retain a connection to the metallic, covalent, and
ionic limits of the VAK triangle. The decoder contains three
components associated with these limits. A learned weight combines
their outputs, with weights guided toward the barycentric
coordinates obtained from the mean electronegativity
$\bar{\chi}$ and the absolute difference $\Delta\chi$.
The weights can depart from this reference during training.
The VAK construction therefore guides the prediction of molecular
properties without assigning fixed bond labels to individual
molecules.

Reference molecular constants cover only a small part of the
possible combinations of atoms. We therefore train the model in
two stages. We first pretrain it on Gaussian-process estimates of
dipole moments for $4{,}851$ neutral
diatomics~\cite{Elhalawani2026}, with estimates weighted by their
inverse variance subject to an uncertainty floor, giving us access to a wider variety of atoms within diatomic molecules. We then
fine-tune it using reference molecular constants. The labeled
dataset contains $290$ neutral molecules, with dipole moments for
all of them ($157$ experimental and $133$ calculated), and
$D_e$, $R_e$, and $\omega_e$ for $110$, $130$, and $130$ molecules,
respectively. Of these molecules, $262$ form the training set
and $28$ are reserved for a fixed held-out test.

The predictions are combined over an ensemble of $50$ members
derived from five independently pretrained models. The decoder
supplies the dipole predictions. For the other three molecular
constants, we use Gaussian processes fitted to the
exchange-invariant features
$[\mathbf{z}_{\mathrm{sym}},|\mathbf{z}_{\mathrm{diff}}|]$,
where the absolute value is taken component-wise. These Gaussian processes
are fitted within each member's training fold and leave the
learned representation unchanged.

\subsection*{A bonding coordinate beyond electronegativity}
\label{sec:coordinate}

We first examine the symmetric channel, which describes each
molecule through the mean of its two atomic encodings.
Its leading principal component contains $46\%$ of the variance
and correlates strongly with the mean electronegativity, with
$r=-0.98$. The representation therefore retains the
metallic-to-covalent ordering of the VAK triangle. Electronegativity
enters both the atomic inputs and the decoder prior, so this
ordering provides a reference against which to examine the
additional information retained during molecular training.

Independently pretrained models can have different learned internal representations (i.e., latent) 
orientations. For the geometric analysis, we therefore use the
mean representation of the ten ensemble members fine-tuned from
one pretrained model. To identify variation beyond the VAK
description, we fit the symmetric representation as a function
of $(\bar{\chi},\Delta\chi)$ using kernel ridge regression and
write
\begin{equation}
 \mathbf{z}_{\mathrm{sym}}
 =
 \widehat{\mathbf{z}}_{\mathrm{VAK}}(\bar{\chi},\Delta\chi)
 +
 \mathbf{z}_{\perp}.
 \label{eq:vak-residual}
\end{equation}
The regression is fitted on the $262$ molecules of the training set.
We define the residual coordinate $\xi$ as the projection of
$\mathbf{z}_{\perp}$ onto its leading principal direction,
\begin{equation}
 \xi \equiv \mathrm{PC}_1(\mathbf{z}_{\perp}).
 \label{eq:xi}
\end{equation}
The fitted VAK dependence and the residual principal direction
are then applied to the pretraining set without refitting. The residual contains $34\%$ of the variance of the symmetric
representation, and $\xi$ accounts for $33\%$ of this residual
variance. The corresponding fraction ranges from $31\%$ to
$37\%$ across the five pretrained models. On the training
set, the correlations of $\xi$ with both VAK coordinates
are $|r|<0.01$. The coordinate represents a leading
contribution to the variation left after the fitted VAK
dependence has been removed.

The shared atomic encoder allows this residual coordinate to
be resolved into atomic contributions given that $\mathbf{z}_{\mathrm{sym}}$ is the mean of the two atomic
encodings and the principal-component projection is linear,
\begin{equation}
 \xi(AB)
 =
 \tfrac{1}{2}\big[e(A)+e(B)\big]
 +
 g(\bar{\chi},\Delta\chi).
 \label{eq:xi-decomposition}
\end{equation}
Here $e(X)$ is the projection of the encoding of element $X$
onto the residual principal direction. The function $g$ is the
negative projection of the fitted VAK contribution, including
the centering constant. Both terms follow the normalization
chosen for $\xi$. This decomposition follows from the
construction of the representation. Its chemical content lies
in the ordering of the atomic contributions learned from the
molecular constants. The atomic contribution \(e(X)\) provides a transferable, element-specific descriptor of molecular bonding, complementary to electronegativity. Its dependence on valence occupancy and shell structure suggests that it captures chemical information not represented by the electronegativity.

\begin{figure*}[t]
  \centering
  \includegraphics[width=\textwidth]{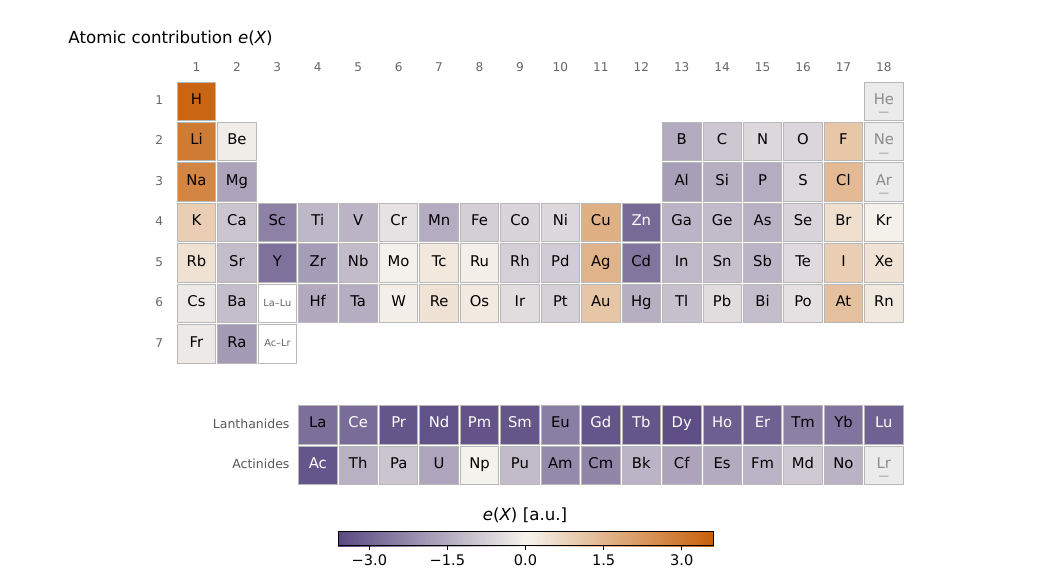}
  \caption{\label{fig:si-atomic-contributions}
  Atomic contributions to the learned bonding coordinate.
  Colors show $e(X)$ for the $99$ elements in the
  pretraining set, obtained by projecting their atomic encodings
  onto the residual principal direction
  [Eq.~\eqref{eq:xi-decomposition}].
  Grey cells with dashes mark elements without a Pauling electronegativity, which are therefore absent from the pretraining set. The values describe the atomic term in the
  decomposition of $\xi$; the residual coordinate also
  contains the VAK-dependent correction
  $g(\bar{\chi},\Delta\chi)$.}
\end{figure*}

Figure~\ref{fig:si-atomic-contributions} shows $e(X)$ across the periodic table. The light alkali metals, the coinage metals, and the halogens have particularly large contributions. Diatomic molecules containing these elements can show substantial departures from the trends in ionic character and dipole moment expected from electronegativity differences alone. Li and Na have an $ns^1$ valence configuration, while Cu, Ag, and Au have $(n-1)d^{10}ns^1$ configurations. The light alkali and coinage metals lie close together in the learned ordering despite belonging to different groups and occupying different regions of the electronegativity scale. The halogens, which are one electron short of a filled valence shell, also have large contributions even though they lie at the opposite end of the electronegativity scale from the alkali metals. The smallest contributions occur for Zn and Cd, with filled $d^{10}s^2$ valence shells, and for the lanthanides. Within the alkali metals, $e(X)$ decreases with increasing period, placing Cs and Fr closer to the middle of the distribution than Li and Na. The ordering therefore reflects both valence occupancy and shell structure, rather than a count of valence electrons alone.

These atomic quantities are supplied to the model as inputs.
The learned ordering shows how they are organized when the
representation is trained to describe molecular bonding. Symbolic regressions~\cite{pysr_paper,Munoz_2025} using different
combinations of the atomic quantities reproduce $\xi$ with
held-out $R^2$ values between $0.70$ and $0.78$, and all of the
selected expressions contain the valence-electron count.
These approximations support the association with valence
occupancy, while the correlations among the atomic inputs
prevent a unique attribution to individual quantities.
The expressions and fitting procedure are given in the
Supplementary Information.

%The residual coordinate extends the VAK description as shown
%in Fig.~\ref{fig:plane}. Each molecule is placed on the triangular
%floor according to $(\bar{\chi},\Delta\chi)$ and displaced
%vertically along $\xi$. The floor retains the metallic,
%covalent, and ionic reading of the VAK triangle. The additional
%coordinate separates molecules with similar electronegativity
%coordinates according to the remaining variation in their
%atomic representations.

\begin{figure*}[t]
  \centering
  \includegraphics[width=\textwidth]{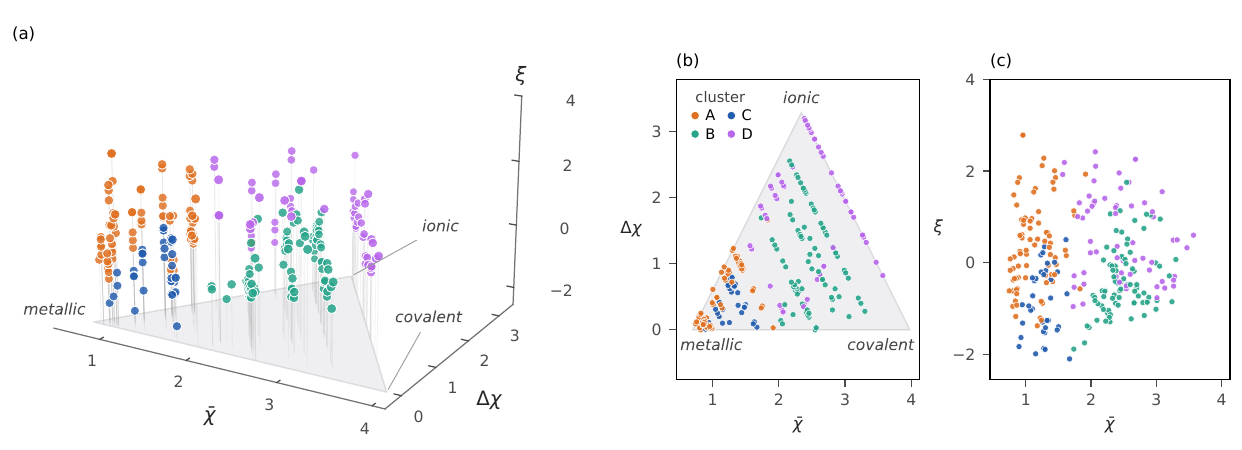}
  \caption{\label{fig:plane}
  The VAK triangle extended by the residual coordinate $\xi$.
  (a) The $262$ molecules of the training set are placed on the VAK
  triangle by their mean electronegativity $\bar{\chi}$ and
  absolute electronegativity difference $\Delta\chi$, and
  displaced vertically along $\xi$. Thin vertical lines
  connect each molecule to its position on the triangle.
  (b) Projection onto the VAK triangle.
  (c) Projection onto the $(\bar{\chi},\xi)$ plane.
  The coordinate $\xi$ is the leading principal component of
  the symmetric representation after subtraction of its fitted
  dependence on the VAK coordinates. It is shown in standardized
  units. Colors indicate the four clusters (A--D) identified by
unsupervised clustering of the learned representation (see Methods).}
\end{figure*}

The residual coordinate $\xi$ extends the VAK triangle into a third dimension, as shown in Fig.~\ref{fig:plane}. Each molecule is positioned in the triangular plane by $(\bar{\chi},\Delta\chi)$ and displaced vertically by $\xi$. The plane retains the conventional metallic, covalent, and ionic regions of the VAK triangle, while the vertical coordinate distinguishes molecules with similar electronegativity coordinates by capturing variation beyond the fitted VAK dependence. In this three-dimensional representation, we identify clusters that group molecules by similarities captured jointly by the electronegativity coordinates and the learned residual. The atomic contribution $e(X)$ orders the elements within the learned representation. For a diatomic molecule $AB$, the residual coordinate $\xi(AB)$ combines the contributions of its two constituent atoms with the VAK-dependent correction to quantify the molecule's departure from the fitted electronegativity trend. We next examine how molecular bonding properties vary across this representation. In the following analysis, we identify four groups of diatomics with distinct bonding characteristics within this learned representation.

\subsection*{Chemical information in the learned representation}
\label{sec:encodes}

The atomic contributions associate the residual coordinate with
valence occupancy and shell structure. We now examine how the
learned representation relates to molecular bonding. We first
test the information it retains about formal bond order, then
compare molecules with different degrees of charge separation
and bond strength.

\begin{figure}[t]
  \centering
  \includegraphics[width=0.95\linewidth]{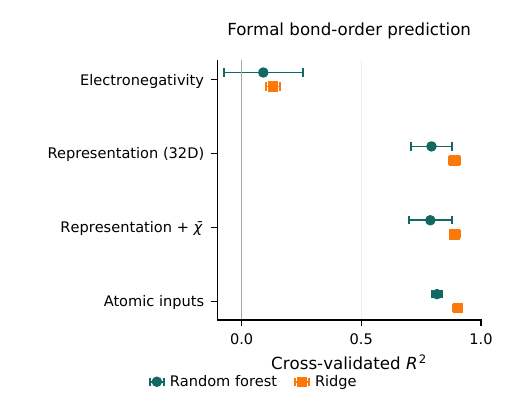}
  \caption{\label{fig:si-bondorder}
  Formal bond-order information in the learned representation.
  Cross-validated predictions are compared for $64$ distinct
  main-group molecules with high-confidence formal bond-order
  assignments, using random-forest and ridge regressions.
  Points show the mean pooled out-of-fold $R^2$ over different random seeds. The learned-representation
  result uses the full $32$-dimensional representation.}
\end{figure}

Figure~\ref{fig:si-bondorder} compares predictions of formal bond
order for $64$ molecules with high-confidence
assignments. Linear regression on the full $32$-dimensional
representation gives a mean pooled out-of-fold
$R^2\approx0.89$. On the other hand, the corresponding values for the
electronegativity-only baselines range from $0.09$ to $0.13$
across the two regressors. Ridge regression on the complete
atomic inputs gives $R^2\approx0.90$. The learned representation
therefore retains bond-order information present in the atomic
descriptors that is poorly captured by the VAK coordinates
alone.

To connect this quantitative comparison with the chemical
organization of Fig.~\ref{fig:plane}, we examine representative
molecules from four regions of the learned space. The examples
include multiple-bonded main-group molecules (cluster B), alkali halides (cluster D),
weakly bound heterometallic molecules (cluster A), and Zn-containing pairs (cluster C). Four clusters are used to organize the comparison. They divide
a continuous representation and are not interpreted as
distinct bond types. Their construction is described in
Methods, and the molecular properties and membership
information are given in Supplementary
Tables~\ref{tab:clusters} and \ref{tab:cluster-members}.

Alongside the spectroscopic constants, we compare the
effective charge (i.e. Pauling Ionic Character),
\begin{equation}
 q_{\mathrm{eff}}
 =\frac{|\mu|}{e_{\mathrm{el}}R_e},
 \label{eq:qeff}
\end{equation}
where $e_{\mathrm{el}}$ is the elementary charge. This
dimensionless quantity expresses the measured or calculated
dipole as a charge separation over the equilibrium bond
length. We also compare calculated Mulliken bond orders,
which describe the representative molecules and are distinct
from the formal bond-order assignments used in
Fig.~\ref{fig:si-bondorder}. The main-group examples SiO, NO, GeO, GeS, and PN in cluster B
have Mulliken bond orders between $1.83$ and $2.86$.
Among the selected examples, they have the shortest bond
lengths, the largest vibrational frequencies, and the deepest
potential wells. Their effective charges range from $0.03$
to $0.43$, consistent with bonds that combine substantial
shared electron density with varying degrees of polarity.

The alkali halides KCl, NaCl, KBr, RbCl, and RbBr in cluster D
have effective charges between $0.768$ and $0.799$, while
their Mulliken bond orders lie between $0.11$ and $0.23$.
Their large effective charges and small bond orders distinguish
them from the multiple-bonded main-group examples.
Their dissociation energies nevertheless remain large,
consistent with substantial ionic binding.

The examples from clusters A and C are weakly bound.
The alkali--alkaline-earth pairs in cluster A have long
equilibrium bond lengths, small vibrational frequencies,
and dissociation energies between $0.15$ and $0.27$~eV.
Their Mulliken bond orders are small but nonzero, and their
effective charges remain below $0.073$. The Zn-containing
examples in cluster C have still shallower potential wells,
with dissociation energies between $0.046$ and $0.055$~eV.
Their effective charges are below $0.009$, and their
Mulliken bond orders range from $0.02$ to $0.15$.

These comparisons distinguish different aspects of bonding.
Effective charge separates the ionic alkali halides from
the weakly polar metal pairs, while bond order and the
spectroscopic constants distinguish the strongly bound
main-group molecules from the weakly bound species.
The VAK coordinates provide the broad metallic, covalent,
and ionic organization. The extended representation retains
additional atomic information associated with valence and
shell structure, and the bond-order comparison demonstrates
that this information remains accessible for molecular
prediction.

\subsection*{Predictive accuracy, uncertainty, and chemical coverage}
\label{sec:generalization}

We test the molecular information retained by the representation
through predictions of the dipole moment, dissociation energy,
equilibrium bond length, and vibrational frequency.
We first evaluate the dipole predictions on $28$ held-out
molecules whose reference values are excluded from every
stage of training. Because the original Gaussian-process
pretraining set includes reference dipoles from this set,
we reconstruct the upstream Gaussian process without those
molecules and repeat pretraining and fine-tuning.
The resulting predictions give $R^2=0.953$ and an RMSE of
$0.60$~D. The reconstruction and validation procedures are
described in Methods and the Supplementary Information.

\begin{figure*}[t]
  \centering
  \includegraphics[width=\textwidth]{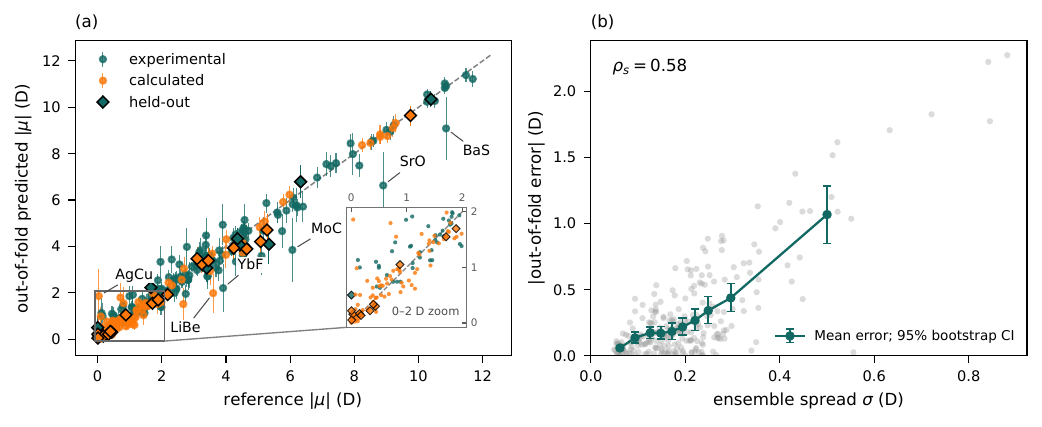}
  \caption{\label{fig:parity}
  Dipole predictions and ensemble uncertainty.
  (a) Out-of-fold predictions against reference dipole
  magnitudes for the original ensemble. Colors distinguish
  experimental and calculated references; black-edged
  diamonds denote the fixed held-out molecules.
  Vertical bars give calibrated $68\%$ prediction intervals,
  and the dashed line denotes equality. The inset expands
  the $0$--$2$~D region. The original pretraining set is
  used throughout this panel; the independent evaluation
  is reported in Table~\ref{tab:cv}.
  (b) Absolute out-of-fold error against ensemble spread.
  Grey points show individual molecules; teal points show
  binned mean errors with $95\%$ bootstrap confidence
  intervals. The Spearman correlation is $\rho_s=0.58$.}
\end{figure*}

For the original ensemble, out-of-fold predictions exclude
each molecule of the training set during fine-tuning, while retaining
the original pretraining set. This evaluation gives
$R^2=0.969$ for the dipole and $0.842$, $0.978$, and $0.964$
for $D_e$, $R_e$, and $\omega_e$, respectively
(Table~\ref{tab:cv}). Figure~\ref{fig:parity}(a) shows the
dipole predictions, including the original ensemble's
fixed held-out set.

\begin{table}[t]
  \centering
  \caption{\label{tab:cv}
  Prediction of ground-state molecular properties.
  The independent dipole test excludes the $28$ held-out
  reference dipoles from both upstream Gaussian-process
  fitting and subsequent training. Out-of-fold results
  exclude molecules during fine-tuning but use the original
  pretraining set. RMSE values use the units given
  for each target.}
    \setlength{\tabcolsep}{9.0pt}

  \begin{tabular}{llrrr}
    \toprule
    Evaluation & Target & $N$ & $R^2$ & RMSE \\
    \midrule
    Held-out
      & $|\mu|$ (D) & 28 & 0.953 & 0.60 \\
    \midrule
    Out-of-fold
      & $|\mu|$ (D) & 262 & 0.969 & 0.485 \\
      & $D_e$ (eV) & 104 & 0.842 & 0.813 \\
      & $R_e$ (\AA) & 123 & 0.978 & 0.091 \\
      & $\omega_e$ (cm$^{-1}$) & 123 & 0.964 & 139.8 \\
    \bottomrule
  \end{tabular}
\end{table}

Direct regressions on the atomic descriptors provide a
comparison on the same labeled folds. The representation
predicts the dipole and vibrational frequency more accurately
than a direct Gaussian process, while the direct model is
more accurate for dissociation energy and slightly more
accurate for bond length. The dipole prediction benefits
from the additional pretraining signal. The representation
therefore provides a common description of molecular
properties.

The ensemble spread also carries information about prediction
error. Larger spreads tend to accompany larger absolute
dipole errors, with Spearman coefficient $\rho_s=0.58$
[Fig.~\ref{fig:parity}(b)]. We calibrate this spread into
prediction intervals using the labeled molecules.
The nominal $68\%$ and $95\%$ intervals achieve mean
coverage of $0.68$ and $0.96$ under repeated calibration
and evaluation splits. On the original ensemble's fixed
held-out set, $23$ of $28$ reference dipoles fall within
the $68\%$ intervals and $26$ within the $95\%$ intervals.
The spectroscopic uncertainties are calibrated in the
same way. Details are given in Methods and the
Supplementary Information.

The uncertainties must be interpreted alongside the
chemical coverage of the data. Of the $262$ molecules
of the training set, $63$ contain fluorine or oxygen, whereas none
pairs a coinage metal with a pnictogen or with a chalcogen
heavier than sulfur. Predictions for combinations such
as AgAs, AuBi, and AgTe therefore extend beyond the
labeled atom pairs. For example, the predicted vibrational
frequency of AgAs has a $68\%$ half-width of approximately
$35\%$ of its value, compared with $10\%$ for PaO.
The candidate predictions and their intervals are given
in Supplementary Table~\ref{tab:candidate-properties}.

Uncertainty and distance within the learned representation
provide complementary indicators for examining sparsely
represented molecules, including radium-, francium-, and
actinide-containing species. We next use the representation to
guide excited-state calculations relevant to optical cycling.

\subsection*{Application to optical-cycling candidate selection}
\label{sec:discovery}

The search for molecules suitable for laser cooling tests
whether the representation can guide calculations of properties
beyond its ground-state training targets. Favorable optical
cycling requires similar ground- and excited-state potentials,
together with a sufficiently strong transition and limited
decay into other electronic states. In established cycling
schemes, the transition often involves an electron localized
mainly on one atom that contributes little to the
bond~\cite{DiRosa2004,FitchTarbutt2021}. Ground-state bonding
can therefore motivate candidate selection, while explicit
excited-state calculations are needed to assess the proposed
transition.

We rank the $4{,}851$ neutral diatomics using a classifier
trained to distinguish $83$ optical-cycling reference species
and well-studied candidates from the remaining molecules,
treated as an unlabeled background. Its inputs are the
ground-state representation, decoder weights, predicted molecular constants,
and ensemble spread. No excited-state potential, transition
dipole, lifetime, or Franck--Condon factor enters the ranking. The
classifier is further described in Methods.

\begin{figure}[t]
  \centering
  \includegraphics[width=\linewidth]{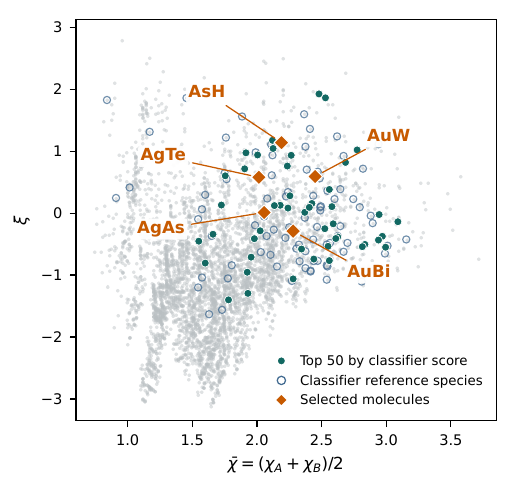}
  \caption{\label{fig:showcase}
  Optical-cycling candidates in the learned bonding
  description. Grey points show the $4{,}851$ molecules
  of the pretraining set in the plane of mean electronegativity
  $\bar{\chi}$ and residual coordinate $\xi$.
  Teal circles mark the $50$ highest-ranked new candidates.
  Open blue circles denote the optical-cycling reference
  species and well-studied candidates used to train the
  classifier. Orange diamonds identify the molecules
  listed in Table~\ref{tab:shortlistFCF}.}
\end{figure}

Figure~\ref{fig:showcase} places the selected molecules in
the $(\bar{\chi},\xi)$ plane. The ranking extends beyond
the alkaline-earth monohalides that motivate many established
cycling schemes. Its Spearman correlation across independent
retrainings is $0.98$--$0.99$ over the full pretraining set,
indicating that the overall ordering is reproducible.
The stability of the top of the ranking is reported in the
Supplementary Information.

We established the shortlist before performing any excited-state calculations. Table \ref{tab:shortlistFCF} lists the five candidates we selected. Among them, AgAs and AuBi are analogues of AsH, which has been studied in the context of laser cooling previously~\cite{Li2022}. Both molecules share the same proposed triplet-to-triplet cooling transition as AsH, $X\,^3\Sigma^-\rightarrow A\,^3\Pi$. AgTe and AuW are natural coinage-metal analogues of TeH and WH \cite{Setzer2001,koseki2006}. The proposed cooling transition in AgTe is the doublet--doublet $X\,^2\Pi\rightarrow A\,^2\Sigma^+$ transition, whereas in AuW it is the sextet--sextet $X\,^6\Pi\rightarrow A\,^6\Delta$ transition. These parallels suggest a connection between the proposed cooling transitions and the single outer $s$ electron shared by hydrogen and the coinage metals. For each candidate, we computed ground- and excited-state potential energy curves using CCSD(T) or internally contracted multireference configuration interaction with the Davidson correction (ic-MRCI+Q), as appropriate for the electronic state. We accounted for scalar relativistic effects through energy-consistent pseudopotentials and omitted spin--orbit coupling. We obtained Franck--Condon factors from the vibrational wavefunctions supported by these potentials and estimated radiative lifetimes from the transition dipole moments and transition frequencies. Further computational details are provided in Methods.

\begin{table}[t]

  \centering
  \caption{\label{tab:shortlistFCF}
  Proposed transitions and calculated radiative properties
  for selected molecules. The AsH values are taken from
  Ref.~\cite{Li2022} and are used as a reference.
  AgAs, AgTe, and AuW are calculated in the
  $\Lambda$--$S$ representation without explicit
  spin--orbit coupling. Lifetimes are estimated for the
  specified decay channels.}
    \setlength{\tabcolsep}{9.0pt}

  \begin{tabular}{llcc}
    \toprule
    Molecule & Proposed transition & $f_{00}$ & Lifetime \\
    \midrule
    AsH
      & $X\,^3\Sigma^-\rightarrow A\,^3\Pi$
      & 0.966 & 914~ns \\
    AgAs
      & $X\,^3\Sigma^-\rightarrow A\,^3\Pi$
      & 0.997 & 13.9~$\mu$s \\
    AuBi
      & $X\,^3\Sigma^-\rightarrow A\,^3\Pi$
      & 0.863 & 3.3 $\mu$s \\
    AgTe
      & $X\,^2\Pi\rightarrow A\,^2\Sigma^+$
      & 0.937 & 21~$\mu$s \\
    AuW
      & $X\,^6\Pi\rightarrow A\,^6\Delta$
      & 0.713 & 0.58~ms \\
    \bottomrule
  \end{tabular}
\end{table}

\begin{figure*}[t]
  \centering
  \includegraphics[width=\textwidth]{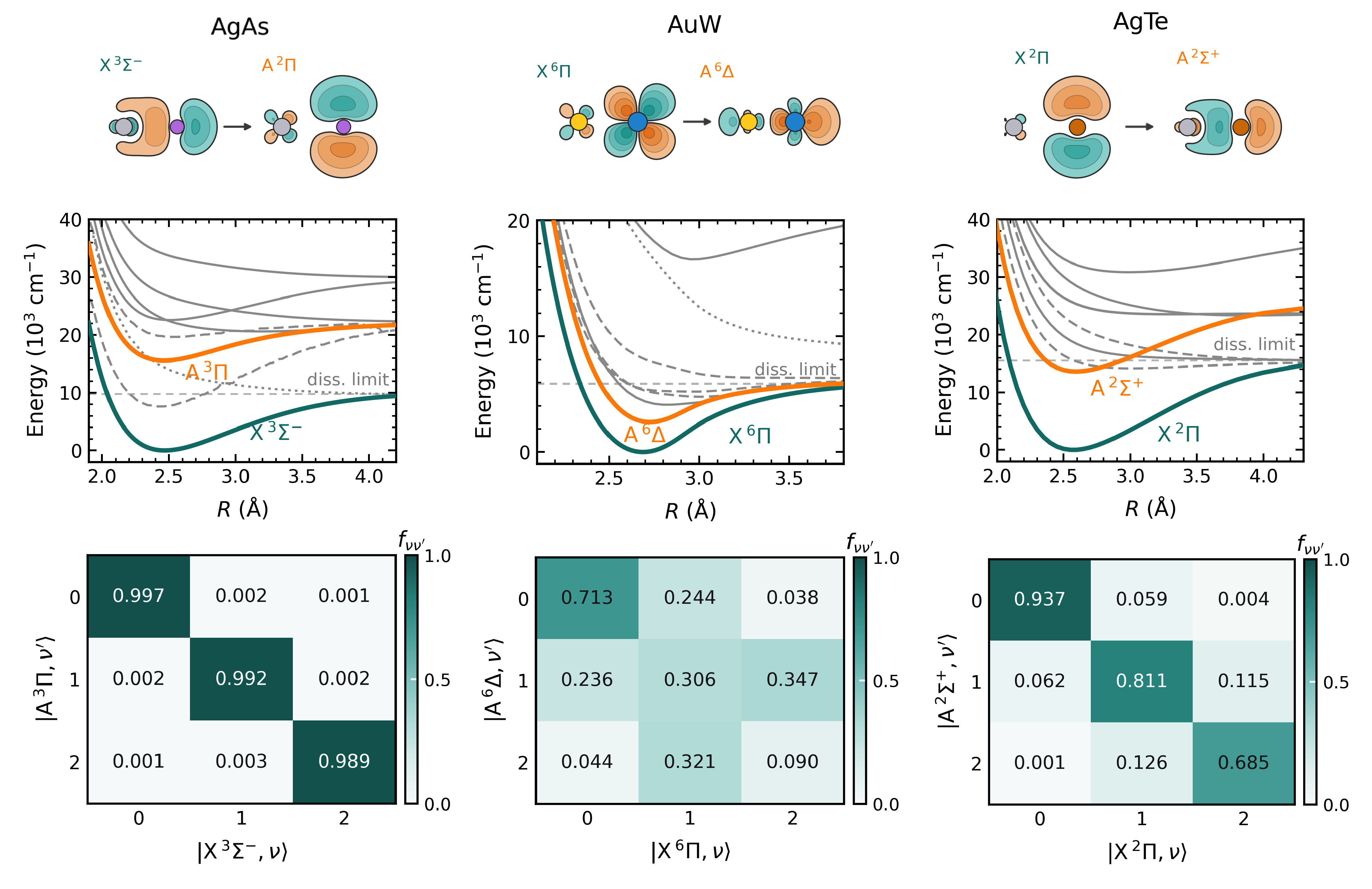}
  \caption{\label{fig:FCF}
  Electronic structure and vibrational branching of the
  calculated candidates. Each column corresponds to one
  molecule. The top row shows the valence orbitals involved
  in the proposed transition. The middle row shows
  $\Lambda$--$S$ potential-energy curves, with the two
  states of the proposed cycle colored and other
  calculated states shown in grey. Dashed and dotted
  curves distinguish spin symmetries. Energies are
  relative to the ground-state minimum.
  The bottom row shows Franck--Condon factors
  $f_{\nu\nu'}$ for $\nu,\nu'=0,1,2$, with lower-state
  vibrational levels along the horizontal axis and
  upper-state levels along the vertical axis.
  All matrices use the same color scale.
  Spin--orbit coupling is not included.}
\end{figure*}

Figure~\ref{fig:FCF} connects the calculated transition
orbitals, potential-energy curves, and vibrational overlaps. Given that AsH, AgAs, and AuBi
present the same cooling transition, and AsH has been studied before \cite{Li2022}, we present the potential curves for AgAs as the example for the triplet-triplet transition.
For AgAs, the proposed
$X\,^3\Sigma^-\rightarrow A\,^3\Pi$ transition has
$f_{00}=0.9967$. The equilibrium bond lengths of the two
states differ by approximately $0.005$~\AA, and the
transition involves predominantly As-centered valence
orbitals. The calculated vibrational overlap therefore
supports the analogy with AsH. The estimated lifetime
of $13.9~\mu$s nevertheless indicates that favorable
vibrational branching is accompanied by a relatively
slow radiative transition.

For AgTe, the proposed $X\,^2\Pi\rightarrow A\,^2\Sigma^+$ transition has a Franck--Condon factor of $f_{00}=0.937$ and an estimated radiative lifetime of $21~\mu\mathrm{s}$. The orbitals involved in the transition are predominantly localized on Te, and the Franck--Condon matrix is less diagonal than that of AgAs. Spin--orbit splitting of the $X\,^2\Pi$ state may allow spontaneous emission into both ground-state $\Omega$ components, as illustrated by TeH~\cite{Setzer2001}. Assessing the closure of the optical cycle therefore requires $\Omega$-resolved potential energy curves and transition strengths.

AuW illustrates a different limitation. Its proposed $X\,^6\Pi\rightarrow A\,^6\Delta$ transition has a Franck--Condon factor of $f_{00}=0.713$, and the orbitals involved in the transition are less localized on a single atom than in AgAs. The electronic states are separated by $2597~\mathrm{cm}^{-1}$, and the estimated radiative lifetime is $0.58~\mathrm{ms}$. Spin--orbit coupling may mix electronic states of the same $\Omega$ and alter the transition strengths inferred from the $\Lambda$--$S$ description. AuW thus ranks highly in the ground-state screening, but its calculated transition has a less favorable vibrational overlap for optical cycling than AgAs.

The calculations establish the opportunity and
the limits of selection from ground-state bonding.
The representation directs attention to chemical
families beyond the reference species, and AgAs
provides a highly diagonal calculated transition.
Transition strengths, spin--orbit mixing, rotational
branching, and electronic leakage remain necessary
parts of an optical-cycling assessment.

\section*{Discussion}
\label{sec:discussion}

The VAK triangle captures the leading organization of diatomic
bonding, but the two electronegativity variables do not contain
all the information supplied by the constituent atoms.
The representation we find recovers the metallic-to-covalent
electronegativity ordering and adds a residual coordinate
$\xi$ that separates molecules with similar VAK coordinates
and different valence and shell structures. The learned contribution $e(X)$ is shared across molecules containing element $X$, and its variation across the periodic table is consistent with trends in valence occupancy and shell structure. Together, the three coordinates organize the diatomics studied here into four clusters, extending the conventional three-region VAK classification.

We tested the learned representation as a screening tool for laser cooling, which depends on the properties of both ground and excited electronic states. Using a representation learned from ground-state information, we ranked candidate diatomics by their proximity to previously studied laser-cooling candidates. The ranking required only the properties of the constituent atoms, and we fixed the shortlist before performing any excited-state calculations. Those calculations identified candidates with favorable Franck--Condon factors and revealed limitations for others. The representation thus helps prioritize molecules for the more demanding calculations needed to assess optical cycling. The ability to infer molecular similarities from atomic properties echoes the early work of Mecke, Clark, and Morse on universal relationships among diatomic spectroscopic constants~\cite{Mecke,Clark,Morse}. 

%The same approach could be tested for predicting other molecular properties and ranking candidates for specific experiments. For example, a representation trained on properties relevant to precision measurements could help prioritize molecules for searches for the electron electric dipole moment or for nuclear Schiff moments. This study shows that molecular
%data can be used to construct an interpretable description
%of bonding that guides the search for properties beyond
%its training targets. Calibrated molecular predictions
%complement this description by providing estimates of
%reliability on the evaluated chemistry and identifying
%predictions that require further investigation.

%Laser cooling provides an application beyond the molecular
%properties used to train the representation. By using a simple geometric
%hypothesis, we demonstrate it is possible to identify candidates for laser cooling, which we then examine through \textit{ab initio} calculations.

This study shows that molecular data can yield an interpretable representation of diatomic bonding that guides the search for molecules with properties beyond the training targets. Calibrated ensemble uncertainties track where the model is interpolating versus extrapolating and identify poorly covered regions where new data is most needed. With training targets tailored to precision measurements, the same approach could help prioritize molecules for searches for the electron electric dipole moment or for nuclear Schiff moments.

%This study shows that molecular
%data can be used to construct an interpretable description
%of bonding that guides the search for properties beyond
%its training targets. Calibrated molecular predictions
%complement this description by providing estimates of
%reliability on the evaluated chemistry and identifying
%predictions that require further investigation.

%The inclusion of excited states remains a natural
%extension, allowing the representation to describe
%how excitation changes the bond and to address more
%directly the properties needed for optical cycling.

Additionally, extending such representations to more
complicated molecular structures is a promising avenue
for the prediction of molecular properties and the
interpretation of bonding. The atomic contributions
provide a starting point, while the roles of geometry,
coordination, and the local chemical environment remain
to be established beyond diatomic molecules.

%%%%%%%%%%%%%%%%%%%%%%%%%%%%%%%%%%%%%%%%%%%%%%%%%%%%%%%%%%%%%%%%%%%%%%%
\section*{Methods}

\subsection*{Data}
%TODO: Should we cite/give the sources for each one of the experimental data points?

The pretraining set contains Gaussian-process estimates of the dipole
moments of $4{,}851$ neutral diatomics formed from $99$
elements~\cite{Elhalawani2026}. Each estimate is weighted by
\begin{equation}
 w_i=\frac{1}{\max(\sigma_i,0.2~\mathrm{D})^2},
\end{equation}
where $\sigma_i$ is the error estimate supplied by the Gaussian
process.

\subsection*{Model and training}
The model is a symmetry-constrained $\beta$-variational autoencoder
($\beta$-VAE)~\cite{Kingma2014,Higgins2017} whose decoder predicts molecular
constants instead of reconstructing the atomic inputs. A shared atomic encoder
applies the same multilayer network to each atom. The spectroscopic
constants are predicted jointly, and the dipole prediction is
conditioned on them. The complete decoder contains fewer
than $8{,}000$ parameters. The atomic inputs and
the full architecture are given in the Supplementary Information.

The training loss is
\begin{equation}
 \mathcal{L}=\mathcal{L}_{\mu}+\beta\,\mathcal{L}_{\mathrm{KL}}
 +\lambda_{s}\mathcal{L}_{s}+\lambda_{g}\mathcal{L}_{g}.
\end{equation}
Here $\mathcal{L}_{\mu}$ is the mean squared error of the dipole, and
$\mathcal{L}_{\mathrm{KL}}$ is the Kullback--Leibler divergence of the
two channels from a standard normal distribution. The term
$\mathcal{L}_{s}$ is the mean squared error of the standardized $D_e$,
$R_e$, $\omega_e$, and $\omega_e\sqrt{m_rR_e}$, where $m_r$ is the
reduced mass, evaluated on the available values only. The term
$\mathcal{L}_{g}$ is the Kullback--Leibler divergence between the
barycentric weights calculated from $(\bar{\chi},\Delta\chi)$ and the
gate weights.

The representation is an ensemble of $50$ random train--test
resamplings with a test fraction of $0.15$. Five pretrained models are
fitted from different random initializations, and ensemble member $m$,
indexed from $0$ to $49$, is fine-tuned from pretrained model
$m \bmod 5$ on $262$ of the $290$ labeled molecules.

Both stages use the AdamW optimizer~\cite{Loshchilov2019} with a
constant learning rate, a weight decay of $4.16\times10^{-7}$, and the
gradient norm clipped at $1$. The weight $\beta$ increases linearly from
zero to $2.57\times10^{-4}$ and then remains constant. Pretraining uses
a random $90\%$ of the pretraining set for training and the remaining
$10\%$ for validation, with $\mathcal{L}_{\mu}$ weighted by $w_i$. Its
settings are $\lambda_s=0$, $\lambda_g=0.3$, a learning rate of
$5\times10^{-4}$, batch size $64$, an increase of $\beta$ over $30$
epochs, and at most $200$ epochs. Fine-tuning uses the test part of each
resampling for validation, with $\lambda_s=0.517$, $\lambda_g=0.268$, a
learning rate of $4.69\times10^{-4}$, batch size $16$, an increase of
$\beta$ over $50$ epochs, and at most $5000$ epochs. Once $\beta$ is
constant, training stops when the validation loss has not improved for
$80$ epochs in pretraining or $150$ epochs in fine-tuning, and the
parameters with the lowest validation loss are kept.

\subsection*{Residual coordinate and atomic contributions}
The kernel ridge regression of Eq.~\eqref{eq:vak-residual} uses the
radial-basis-function kernel
$k(\mathbf{v},\mathbf{v}')=\exp[-\gamma\|\mathbf{v}-\mathbf{v}'\|^2]$,
where $\mathbf{v}=(\bar{\chi},\Delta\chi)$, regularization parameter
$\lambda_{\mathrm{KRR}}=0.1$, and kernel parameter $\gamma=1.0$. The
coordinate $\xi$ is scaled to unit variance on the training
set, and its sign is chosen so that it correlates positively with
$|\mu|$ on the same molecules. The atomic contribution $e(X)$ is the
projection of the encoding $\mathbf{h}_X$, averaged over the same
ensemble members as the symmetric representation, onto the residual
principal direction, with the same scaling and sign as $\xi$.

\subsection*{Prediction of constants and validation}
The Gaussian processes that predict $D_e$, $R_e$, and $\omega_e$ from
the representation use the product of a constant and a
radial-basis-function kernel with one length scale per coordinate, plus
a noise term. The inputs are standardized on each member's training
fold, and the kernel parameters are obtained by maximizing the marginal
likelihood.

Each molecule of the training set is scored only by the members that held it
out, and its out-of-fold prediction is the mean over these members.
The dipole is scored in the same way from the decoder output. The $28$
held-out molecules are scored by the mean over all $50$ members.

The Gaussian process that supplies the pretraining set was fitted to
$273$ reference dipole entries~\cite{Elhalawani2026}, which include
$27$ of the $28$ held-out molecules. To remove this overlap, we built
an approximate reconstruction of that Gaussian process following
Ref.~\cite{Elhalawani2026}, with the same atomic descriptors, the
period and group of both atoms, a radial-basis-function kernel, and
both atom orders. Fitted to all entries, it reproduces the published
dipoles with $R^2=0.94$ on the molecules outside its training set. We
then refit it without the held-out molecules and repeat pretraining,
fine-tuning, and calibration with the same settings, so that no
held-out label enters any stage.

\subsection*{Uncertainty calibration}
We calibrate the ensemble spread by split conformal
prediction~\cite{Vovk2005,Lei2018}. For each molecule of the training
set, the score $|\Delta\mu|/\sigma_{\mu}$ is formed from its
out-of-fold residual $\Delta\mu=|\mu|_{\mathrm{pred}}-|\mu|_{\mathrm{ref}}$
and its spread $\sigma_{\mu}$ over the $50$ members. The $68\%$ and
$95\%$ intervals are $\pm c\,\sigma_{\mu}$, where $c$ is the
$\lceil (n+1)q\rceil$-th smallest of the $n$ scores and $q$ is the
nominal coverage. This gives $c_{68}=1.62$ and $c_{95}=3.19$ for the
dipole. For $D_e$, $R_e$, and $\omega_e$, the spread combines the
posterior variance of the Gaussian processes with the variance of their
predictions over the members that held the molecule out, and the
multipliers are obtained in the same way. Coverage is estimated by
calibrating on a random half of the training set and counting on the
other half, averaged over $500$ splits.

\subsection*{Clustering}
The partition is built in the exchange-invariant coordinate
$[\mathbf{z}_{\mathrm{sym}},|\mathbf{z}_{\mathrm{diff}}|]$ of the $262$
molecules of the training set. Each of the $50$ members is clustered
separately, in its own coordinates, with $k$-means, and the final
partition into $K=4$ clusters groups molecules by how often they fall in
the same cluster across members. The confidence of a molecule is defined as the
fraction of members that agree with its final cluster. For
a molecule outside the training set, the membership probabilities are
$k$-nearest-neighbor votes ($k=5$) over the partition, averaged over the
members.

\subsection*{Candidate ranking}
We use a voting ensemble of $50$ gradient-boosted decision-tree
classifiers~\cite{Chen2016}, which take as input the representation
$[\mathbf{z}_{\mathrm{sym}},|\mathbf{z}_{\mathrm{diff}}|]$, the cluster
votes of the ensemble members, $\bar{\chi}$, and $\Delta\chi$. Each classifier
uses $300$ trees with a maximum depth of $6$ and a learning rate of
$0.05$. The final score of a molecule is the average over the $50$
classifiers, and the $50$ candidates are the highest-scoring molecules
that are neither reference species nor in the training set.

\subsection*{Electronic-structure calculations}
All calculations are performed with the MOLPRO quantum-chemistry
package~\cite{Werner2020}. We use the Stuttgart pseudopotentials with
the corresponding aug-cc-pVQZ-PP basis sets, and the molecular orbitals
are obtained from an initial multiconfiguration self-consistent-field
calculation. For each candidate, we include the potential-energy curves
of all atomic dissociation channels up to one channel above the one
containing the diagonal excited state. The vibrational states are
computed with the Colbert--Miller discrete variable
representation~\cite{ColbertMiller1992}. The pseudopotentials and
active spaces of each molecule are given in the Supplementary
Information.

%%%%%%%%%%%%%%%%%%%%%%%%%%%%%%%%%%%%%%%%%%%%%%%%%%%%%%%%%%%%%%%%%%%%%%%
\begin{acknowledgments}
% We thank Arian Jadbabaie for valuable discussions and insights. Computational resources were provided by SubMIT at MIT Physics and by the National Science Foundation under Cooperative Agreement PHY-2525568 (The NSF AI Institute for Artificial Intelligence and Fundamental Interactions, http://iaifi.org/). J.P.-R. and M. L. acknowledge the support from the U.S. National Science Foundation through CAREER Award No. PHY-2440808.
We thank Arian Jadbabaie for valuable discussions and insights. Computational resources were provided by SubMIT at MIT Physics and by the National Science Foundation under Cooperative Agreement PHY-2525568 (The NSF AI Institute for Artificial Intelligence and Fundamental Interactions, http://iaifi.org/). J.P.-R. and M. L. acknowledge the support from the U.S. National Science Foundation through CAREER Award No. PHY-2440808.  J.M.M.  and R.F.G.R. acknowledge support from the MIT Generative AI Impact Consortium (MGAIC), and the U.S. Department of Energy, Office of Science, under Award Number DE-SC0026800 (Genesis Mission).
\end{acknowledgments}

\section*{Data and code availability}
The labeled dataset, the pretraining set, the trained ensemble, and the code
that produces every number, table, and figure in this paper are available upon request.

\newpage

\bibliography{main}
\newpage
\appendix

\setcounter{figure}{0}
\setcounter{table}{0}
\renewcommand{\figurename}{Supplementary Fig.}
\renewcommand{\tablename}{Supplementary Table}
\renewcommand{\thetable}{\arabic{table}}
\section{Model architecture and training}
\label{app:architecture}
\begin{figure*}[t]
  \centering
  \includegraphics[width=\textwidth]{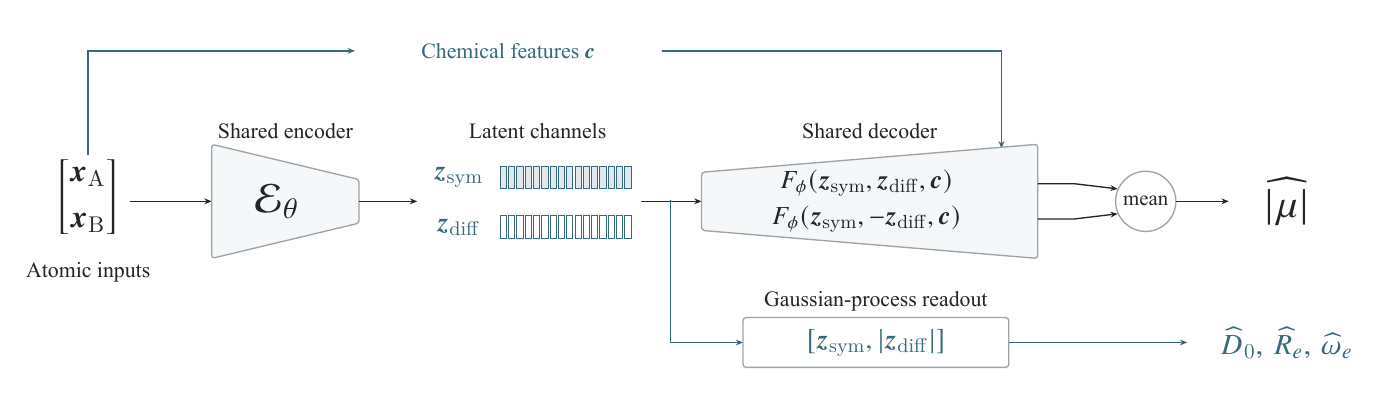}
  \caption{\label{fig:arch}
  Learning a representation of diatomic bonding from atomic
  properties. The shared encoder maps the properties of each atom
  into an atomic representation. Their mean and difference define
  symmetric and antisymmetric molecular channels. The decoder
  combines metallic, covalent, and ionic components through a
  learned gate guided by the Van Arkel--Ketelaar triangle.
  Averaging the decoder output over both signs of the antisymmetric
  channel makes the molecular predictions invariant under exchange
  of the atoms.}
\end{figure*}
The encoder receives the ionization potential, electron affinity, and
electronegativity of the atom, its group and period, its number of
valence and unpaired electrons, and its first atomic excitation energy.
The ionization potential, electron affinity, and electronegativity are
mapped onto a standard normal distribution according to their ranks
among the atoms of the training set, so that the three quantities
share a common scale. The group, the period, and the number of valence
electrons take integer values, and each value is represented by a short
vector learned during training. The number of unpaired electrons and the
excitation energy enter divided by $7$ and by $20$~eV, respectively. The
unscaled ionization potentials, electron affinities, and
electronegativities enter the four chemical quantities defined below and
the barycentric weights of the gate.

The representation has a
$16$-dimensional symmetric channel $\mathbf{z}_{\mathrm{sym}}$ and a $16$-dimensional
antisymmetric channel $\mathbf{z}_{\mathrm{diff}}$, for $32$ dimensions in total. Using the posterior-mean atomic representations $\mathbf{h}_A$ and $\mathbf{h}_B$
defined in Sec.~\ref{sec:plane}, the two channels are
$\mathbf{z}_{\mathrm{sym}}=\tfrac{1}{2}(\mathbf{h}_A+\mathbf{h}_B)$ and
$\mathbf{z}_{\mathrm{diff}}=\mathbf{h}_A-\mathbf{h}_B$.
Exchanging the atoms leaves $\mathbf{z}_{\mathrm{sym}}$ unchanged and
reverses the sign of $\mathbf{z}_{\mathrm{diff}}$. The gate and the decoder are therefore evaluated
on both signs and averaged,
\begin{equation}
 f(\mathbf{z}_{\mathrm{sym}},\mathbf{z}_{\mathrm{diff}})
 \to
 \tfrac{1}{2}\left[f(\mathbf{z}_{\mathrm{sym}},\mathbf{z}_{\mathrm{diff}})+f(\mathbf{z}_{\mathrm{sym}},-\mathbf{z}_{\mathrm{diff}})\right],
\end{equation}
which is exactly invariant under the exchange and gives identical predictions for
$AB$ and $BA$. This average keeps the relative signs between the components of
$\mathbf{z}_{\mathrm{diff}}$, that is, whether the same atom dominates along different
directions of the antisymmetric channel, which squaring each component would
discard (Appendix~\ref{app:validation}).

A matched model that receives the raw signed antisymmetric channel does not satisfy
$AB=BA$. Its dipole changes by $0.7$~D at the median and by as much as $4.0$~D under
exchange of the two atoms. This comparison shows the importance of imposing permutation symmetry
directly.

The four chemical quantities calculated in the forward pass are
\begin{align}
 I_{\mathrm{HS}}
 &=16\Delta\chi+3.5(\Delta\chi)^2,\\
 F_{\mathrm{IP},1}
 &=|\mathrm{IP}_A-\mathrm{IP}_B|^{1/4},\\
 F_{\mathrm{IP},2}
 &=\frac{1}{\mathrm{IP}_A\mathrm{IP}_B},\\
 F_{\mathrm{EA}}
 &=\mathrm{EA}_A+\mathrm{EA}_B.
\end{align}
They are permutation invariant and are concatenated to the decoder input.  The first is the ionic character as defined by Hannay and Smith~\cite{HannaySmyth1946}, while the other three quantities relate to the molecule's electrostatic properties.

The learned gate stays close to the electronegativity prior. Over the labeled set, 
the mean total-variation distance between the two is $0.020$, and their leading
component agrees for $98.5\%$ of the molecules ($97.1\%$ over the pretraining
set). Pinning the gate to the prior at
inference lowers the out-of-fold $R^2$ by $0.020$ for the dipole and by $0.058$ for
$\omega_e$, and a uniform gate collapses the dipole to $R^2=0.50$. The molecules
where the gate moves furthest from the prior are HF, BaS, AuF, CuO, TiS, CuS, FeC,
AgI, PbO, and TlI. That is, hydrides and coinage- or heavy-metal halides and chalcogenides,
the cases in which the electronegativity difference alone is a poor guide to the
bonding character.

The encoder has three residual blocks of width $64$ with dropout $0.2$, and the
decoder uses dropout $0.144$.

\begin{figure}[t]
  \includegraphics[width=\linewidth]{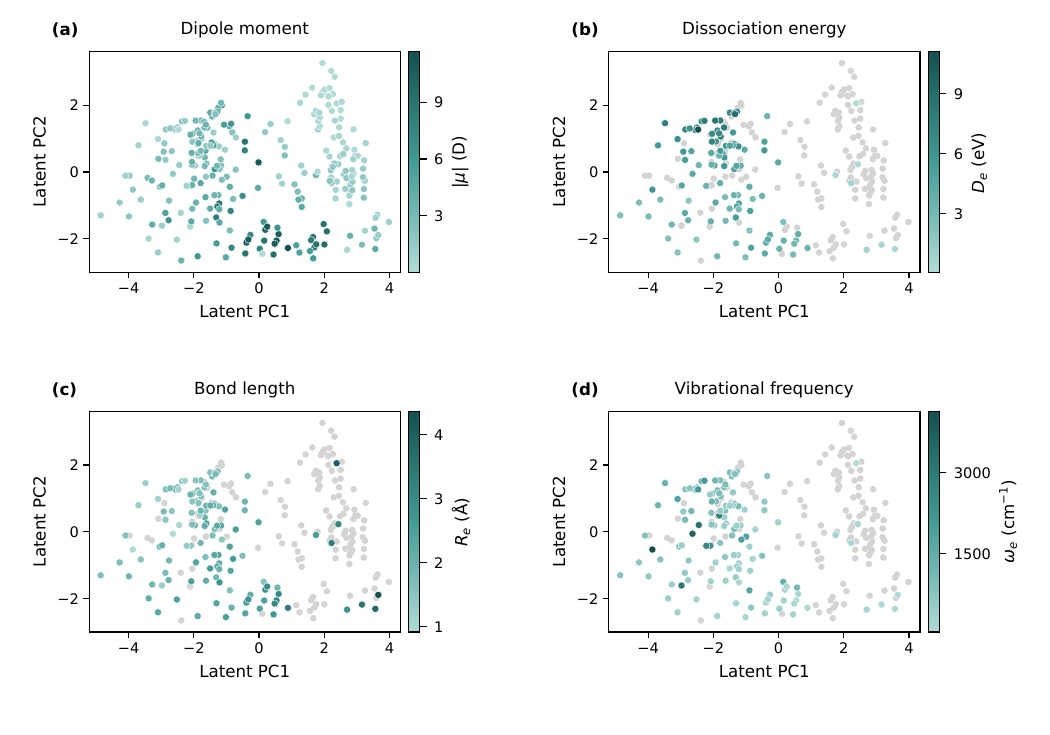}
  \caption{\label{fig:si-decode}
  Molecular constants across the leading plane of the representation. The points
  are colored by (a) dipole moment $|\mu|$, (b) dissociation energy $D_e$,
  (c) equilibrium bond length $R_e$, and (d) vibrational frequency $\omega_e$.
  Grey points do not have a value for the quantity shown in the corresponding
  panel.}
\end{figure}

\section{Validation tests}
\label{app:validation}

The main text reports out-of-fold validation and a fixed held-out test.
Both evaluations exclude the scored molecules during fine-tuning, while their
atom pairs remain in the Gaussian-process pretraining set. The reference
dipoles used to fit the upstream Gaussian process overlap the labeled dataset,
so these evaluations are not fully independent of the reference dipole data.

In the out-of-fold test, the held molecule is removed during fine-tuning, but its
atom pair can remain in the pretraining set. This test measures interpolation
within the chemical range represented during pretraining.

The fixed held-out set contains $28$ molecules excluded from the fine-tuning set of
the ensemble. Their atom pairs remain in the Gaussian-process pretraining
table, as do those of all $262$ fine-tuning molecules, so this set tests the same
setting as the out-of-fold evaluation. The difference is that no member of the
ensemble used these molecules for early stopping either.

\emph{Independence from the pretraining set.} Table~\ref{tab:gp-independent} compares the
held-out dipole with the pretraining set generated by a Gaussian process fitted with and
without the held-out molecules (Methods). The held-out dipole goes from $R^2=0.98$ to
$0.95$ (RMSE $0.39$ to $0.60$~D) and remains accurately predicted when the held-out
dipoles are excluded from pretraining.

\begin{table}[b]
  \caption{\label{tab:gp-independent}
  Dipole on the fixed held-out set, with the pretraining set regenerated from the
  reconstructed Gaussian process fitted with and without the held-out molecules.}
  \begin{ruledtabular}
  \begin{tabular}{lccc}
   Gaussian-process fit & $N$ & $R^2$ & RMSE (D) \\
   \hline
   All entries          & 28 & 0.981 & 0.39 \\
   Held-out removed     & 28 & 0.953 & 0.60 \\
  \end{tabular}
  \end{ruledtabular}
\end{table}

\emph{Prediction of the spectroscopic constants.} The decoder's scalar head under-uses the
representation for $R_e$ and $\omega_e$. A Gaussian process with an
automatic-relevance-determination kernel, fitted to a member's
$[\mathbf{z}_{\mathrm{sym}},|\mathbf{z}_{\mathrm{diff}}|]$ on that member's training fold, predicts the
out-of-fold molecules with $R^2=0.98$ and $0.96$, against $0.94$ and $0.94$ from the head
(Table~\ref{tab:baselines}). Here brackets denote concatenation, and
$|\mathbf{z}_{\mathrm{diff}}|$ denotes the componentwise absolute
value of the antisymmetric channel. This Gaussian process is exchange invariant by construction,
requires no further training of the neural representation, and provides the
spectroscopic predictions reported in this work. The head remains the dipole
predictor, where it is the more accurate of the two ($0.969$ against $0.955$).

\emph{Baselines on the same folds.} Table~\ref{tab:baselines} compares the
representation with regressors fitted directly to the per-atom quantities the
network receives, symmetrised as
$\mathbf{x}_A+\mathbf{x}_B$ and
$|\mathbf{x}_A-\mathbf{x}_B|$, where $\mathbf{x}_A$ and
$\mathbf{x}_B$ are the atomic input vectors and the absolute
value is componentwise, on the same $50$ training folds and
scored the same way. The direct Gaussian process is the stronger
per-property regressor for $D_e$ and $R_e$; the representation is stronger for the
dipole, where the pretraining set contributes information the $262$ labels do not
contain, and for $\omega_e$.

\begin{table}[b]
  \caption{\label{tab:baselines}
  Out-of-fold $R^2$ on the ensemble folds. The first three rows are fitted directly
  to the symmetrised atomic quantities; the last two read the trained representation.
  Every molecule is scored only by the fits that held it out.}
  \begin{ruledtabular}
  \begin{tabular}{lcccc}
   Model & $|\mu|$ & $D_e$ & $R_e$ & $\omega_e$ \\
   \hline
   Ridge regression                    & 0.64  & 0.57  & 0.92  & 0.91 \\
   Random forest                       & 0.87  & 0.67  & 0.87  & 0.74 \\
   GP, atomic inputs     & 0.950 & 0.923 & 0.987 & 0.925 \\
   Decoder head                        & 0.969 & 0.860 & 0.937 & 0.935 \\
   GP on the representation & 0.955 & 0.842 & 0.978 & 0.964 \\
  \end{tabular}
  \end{ruledtabular}
\end{table}

\emph{Pretraining.} Without the pretraining stage, with one random initialization
per member on the same folds, the dipole falls to $R^2=0.952$--$0.955$
($\mathrm{RMSE}$ $0.59$--$0.61$~D against $0.485$~D) over three seeds, while the
decoder's constants do not: $D_e$ $0.86$--$0.87$, $R_e$ $0.955$--$0.963$, $\omega_e$
$0.94$. The distilled task is the dipole, and it is the dipole that gains; the
constants come from the fine-tuning labels. With a single pretrained model instead
of five, the dipole is $0.965$--$0.966$ and the constants spread over $D_e$
$0.81$--$0.87$, $R_e$ $0.91$--$0.94$, and $\omega_e$ $0.90$--$0.92$ across three seeds:
Averaging over five pretraining draws removes that dependence. Replacing the sign average of
Appendix~\ref{app:architecture} by the element-wise square of the antisymmetric
channel costs $0.09$ in the out-of-fold $R^2$ of $R_e$ over six seeds.

\section{Robustness of the bonding coordinate}
\label{app:robustness}

The measured and calculated parts of the molecular dataset occupy different parts
of the periodic table. The data source is consequently strongly correlated with
the learned clusters, with Cram\'er's $V=0.886$, and is approximately $99\%$
predictable from electronegativity alone. Predicting the data source from the
representation gives an area under the curve of $0.999$, compared with $0.998$
from the atomic quantities alone. The representation thus does not create this
separation.

Removing the VAK prior lowers the Rand index
value to $0.655$. The calculated subset is reproduced more clearly than the
experimental subset, with adjusted Rand indices of $0.958$ and $0.561$,
respectively.

\begin{figure*}[t]
  \includegraphics[width=\textwidth]{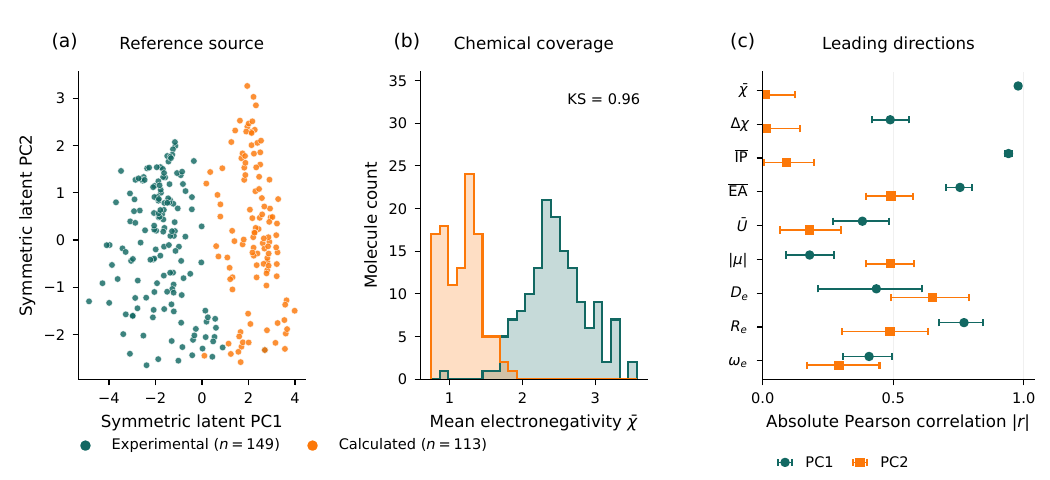}
  \caption{\label{fig:si-confound}
  Effect of the data source on the learned space. (a) The leading plane colored by
  experimental and calculated reference values. (b) Distributions of the mean
  electronegativity for the two subsets. Their Kolmogorov--Smirnov distance is
  $0.96$. (c) Absolute correlations of the leading directions with atomic and
  molecular quantities, with bootstrap $95\%$ intervals.}
\end{figure*}

The number of clusters is not fixed by the data. Over $K=2$ to $8$ the gap
statistic increases monotonically and never satisfies its stopping rule, and the
mean pairwise adjusted Rand index between the member partitions is highest at
$K=2$ ($0.86$), $0.70$ at $K=3$, $0.69$ at $K=4$, and $0.74$ at $K=5$. The silhouette
is maximal at $K=4$ ($0.25$, against $0.23$ at $K=3$ and $0.22$ at $K=5$). We use
$K=4$ because it is the smallest division that places clusters in the three corners
of the VAK triangle and in the region between them. It is also the silhouette
optimum.

\begin{figure*}[t]
  \includegraphics[width=\textwidth]{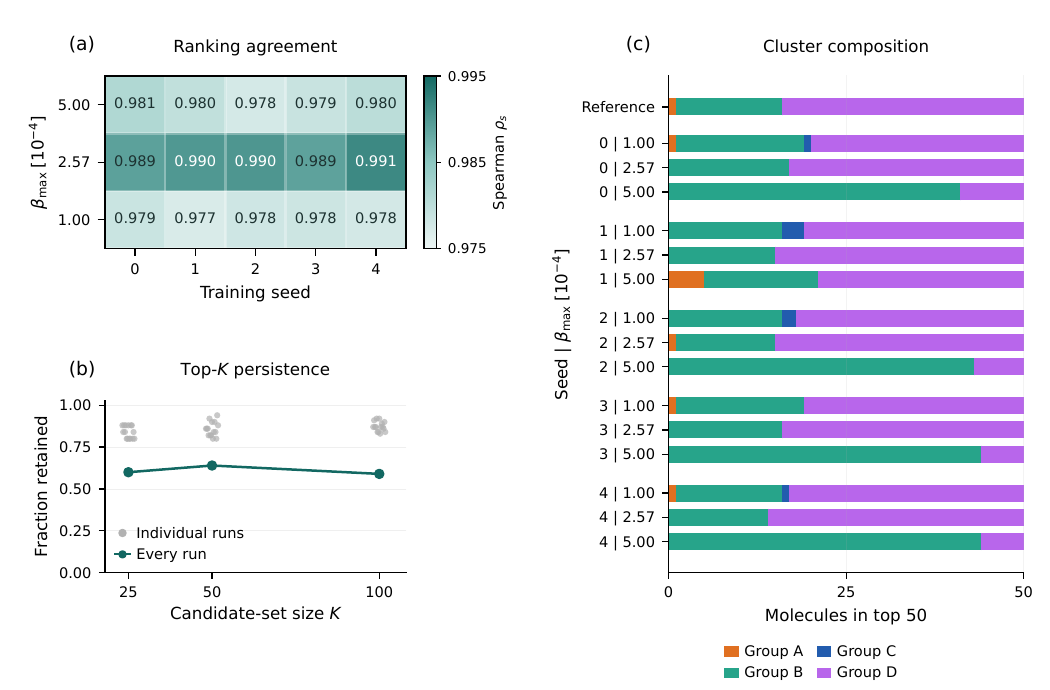}
  \caption{\label{fig:si-robust}
  Stability of the candidate ranking across five training seeds and three values
  of $\beta_{\max}$. (a) Spearman correlation with the full reference ranking.
  (b) Fraction of the reference top-$K$ retained in individual runs (grey) and
  in every run (teal). (c) Cluster composition of each top-$50$ set, including the
  reference run.}
\end{figure*}

Over the $15$ retrainings (five seeds, three values of $\beta_{\max}$), the
electronegativity direction of the representation is stable, with
$|r|=0.94$--$0.98$ against $\bar{\chi}$, and the partition is reproduced with
adjusted Rand indices between $0.77$ and $0.93$ relative to the reference
partition (mean $0.85$); between retrainings the indices lie between $0.71$ and
$1.00$. Across the retrainings, the classifier's
$\mathrm{ROC}$--$\mathrm{AUC}$ ranges from $0.939$ to $0.946$, and the
Spearman correlation of the complete ranking of the pretraining set against the
reference run is $\rho_s=0.98$--$0.99$ [Fig.~\ref{fig:si-robust}(a)].

Formal bond order can be predicted linearly from the full $32$-dimensional
representation with mean pooled out-of-fold $R^2\approx0.89$ over the
high-confidence main-group set of $64$ distinct molecules. The
mean $R^2$ for the electronegativity-only baseline ranges from $0.09$ to $0.13$
across the two regressors, while ridge regression on the complete atomic input
gives approximately $0.90$. The representation therefore carries part of the bond-order information
contained in the atomic identities, but does not define a separate universal
bond-order coordinate.

\section{Quantum-chemistry calculations for the laser-cooling candidates}
\label{app:quantum-chemistry}

The potential-energy curves and molecular orbitals in
the main text were computed using a combination of internally contracted multireference configuration interaction (ic-MRCI) and coupled-cluster theory with single, double, and perturbative triple excitations, CCSD(T). Relativistic effects were included through Stuttgart energy-consistent pseudopotentials. The electronic-structure calculations were performed in the ($\Lambda-S$) representation and did not explicitly include spin–orbit coupling. This representation is sufficient for identifying the orbital character of the relevant electronic states. In particular, the bonding coordinate ($\xi$) characterizes whether the optically active electron occupies a nonbonding orbital, a property of the transition that is not altered by the inclusion of spin–orbit coupling.

\begin{figure*}[t]
 \includegraphics[width=\textwidth]{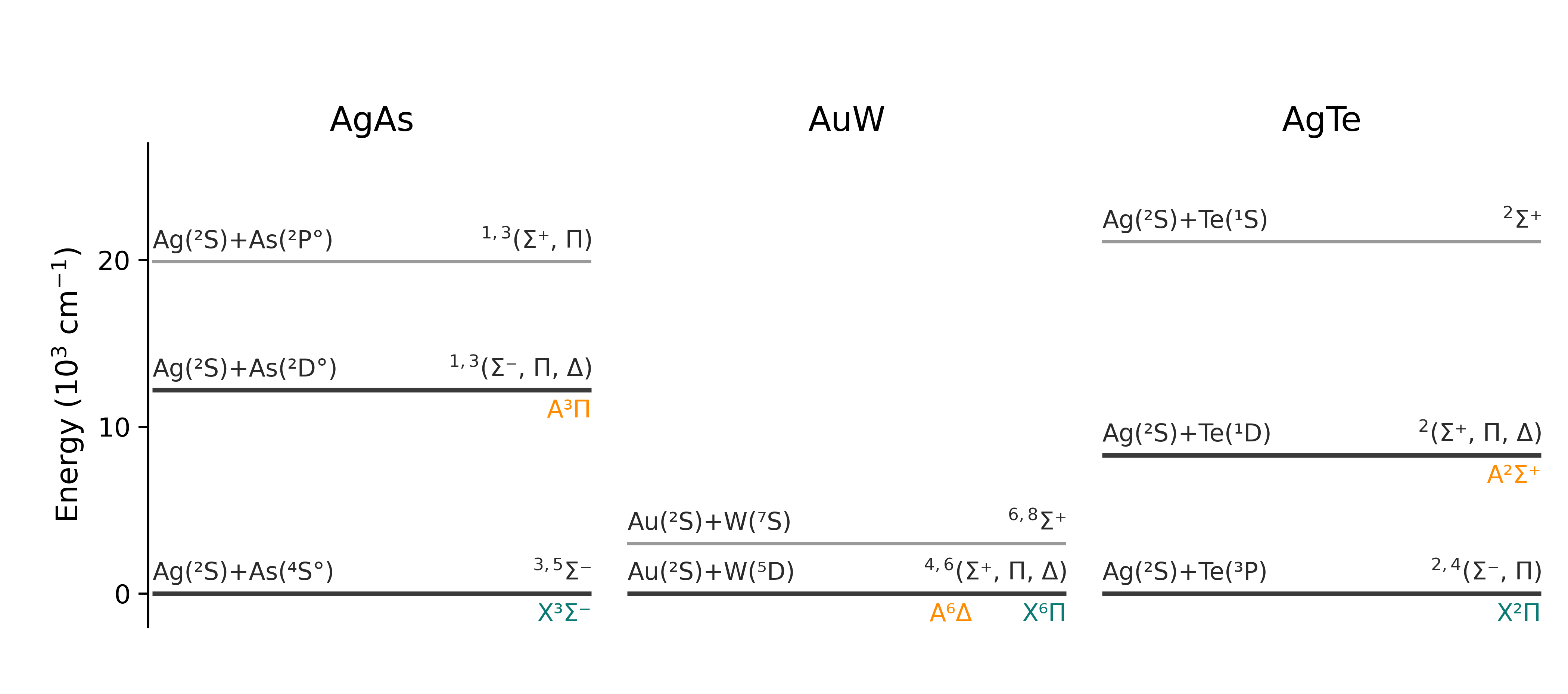}
 \caption{\label{fig:candidates_states}
 Molecular channels and corresponding $\Lambda$–$S$ electronic states for the studied laser cooling candidates.}
\end{figure*}

Figure \ref{fig:candidates_states} illustrates the channels included for each candidate, together with their corresponding molecular states. Furthermore, Table \ref{tab:spectroscopy} tabulates the spectroscopic constants of the states relevant in the cooling transition.

\subsection{AgAs}
For Ag, in both AgAs and AgTe, the inner 28 electrons were replaced by the small-core, scalar relativistic energy-consistent pseudopotential ECP28MDF. The remaining electrons were described using the corresponding augmented correlation-consistent polarized valence quadruple-zeta basis set optimized for the pseudopotential, aug-cc-pVQZ-PP, with the contraction $(15s,12p,11d,4f,3g,2h)/[7s,7p,6d,4f,3g,2h]$. Similarly, for As, the inner 28 electrons were replaced by the corresponding ECP28MDF pseudopotential, while the remaining electrons were described using the aug-cc-pVQZ-PP basis set with the contraction $(29s,12p,13d,3f,2g)/[7s,6p,5d,3f,2g]$.

An initial multiconfiguration self-consistent-field (MCSCF) calculation revealed that the ground state, X$^{3}\Sigma^{-}$, and the first excited triplet state, A$^{3}\Pi$, are predominantly described by single-reference configurations. These states were therefore calculated using the CCSD(T) method. For the remaining electronic states, we employed the internally contracted multireference configuration interaction method with the Davidson correction (ic-MRCI+Q), using the MCSCF orbitals as reference orbitals. The MCSCF active space comprised the $4s$, $4p$, $4d$, $5s$, and $5p$ orbitals of Ag and the $4s$ and $4p$ orbitals of As. Within this active space, the Ag $5s$ and $5p$ and As $4s$ and $4p$ orbitals were included as active orbitals, while the remaining orbitals were kept doubly occupied. All spin multiplicities were treated at the same level of theory.

\subsection{AuW}
For the Au atom, we used the ECP60MDF pseudopotential to replace the 60 innermost electrons, together with the corresponding aug-cc-pVQZ-PP basis set to describe the remaining 19 electrons. The basis set was contracted as $(15s,12p,11d,4f,3g,2h)/[7s,7p,6d,4f,3g,2h]$. The same pseudopotential and basis set were used for W, with the corresponding contraction $(15s,12p,11d,4f,3g,2h)/[7s,7p,6d,4f,3g,2h]$.

In contrast to the AgAs case, several states were found to exhibit significant multireference character. Therefore, all potential-energy curves were calculated using the ic-MRCI+Q method, with molecular orbitals obtained from an initial MCSCF calculation. The MCSCF active space comprised the $5s$, $5p$, $5d$, $6s$, and $6p$ orbitals of Au and the $5s$, $5p$, $5d$, $6s$, and $6p$ orbitals of W. We consider as doubly occupied the $5s$, $5p$, $5d$ orbitals of Au and the $5s$, $5p$ orbitals of W. The active orbitals include the $5d$ orbital of W and the $6s,6p$ orbitals of both atoms.

\subsection{AgTe}
For Te, we used the ECP46MDF pseudopotential to replace the 46 innermost electrons, together with the corresponding aug-cc-pVQZ-PP basis set to describe the remaining six electrons of the $5s$ and $5p$ orbitals. The basis set was contracted as $(15s,23p,13d,3f,2g)/[7s,7p,5d,3f,2g]$.

Similarly to the AgAs case, an initial MCSCF calculation showed that the ground state, X$^{2}\Pi$, and the first excited state, A$^{2}\Sigma^{+}$, are well described by single-reference configurations. These states were therefore calculated at the CCSD(T) level. The remaining electronic states were obtained using the MCSCF/ic-MRCI+Q approach. The MCSCF active space comprised the $4s$, $4p$, $4d$, $5s$, and $5p$ orbitals of Ag and the $5s$, $5p$, and $6s$ orbitals of Te. The $4s$, $4p$, and $4d$ orbitals of Ag together with the $5s$ orbital of Te, were kept doubly occupied, while the remaining orbitals were included as active orbitals.

\begin{figure}[t]
  \includegraphics[width=\linewidth]{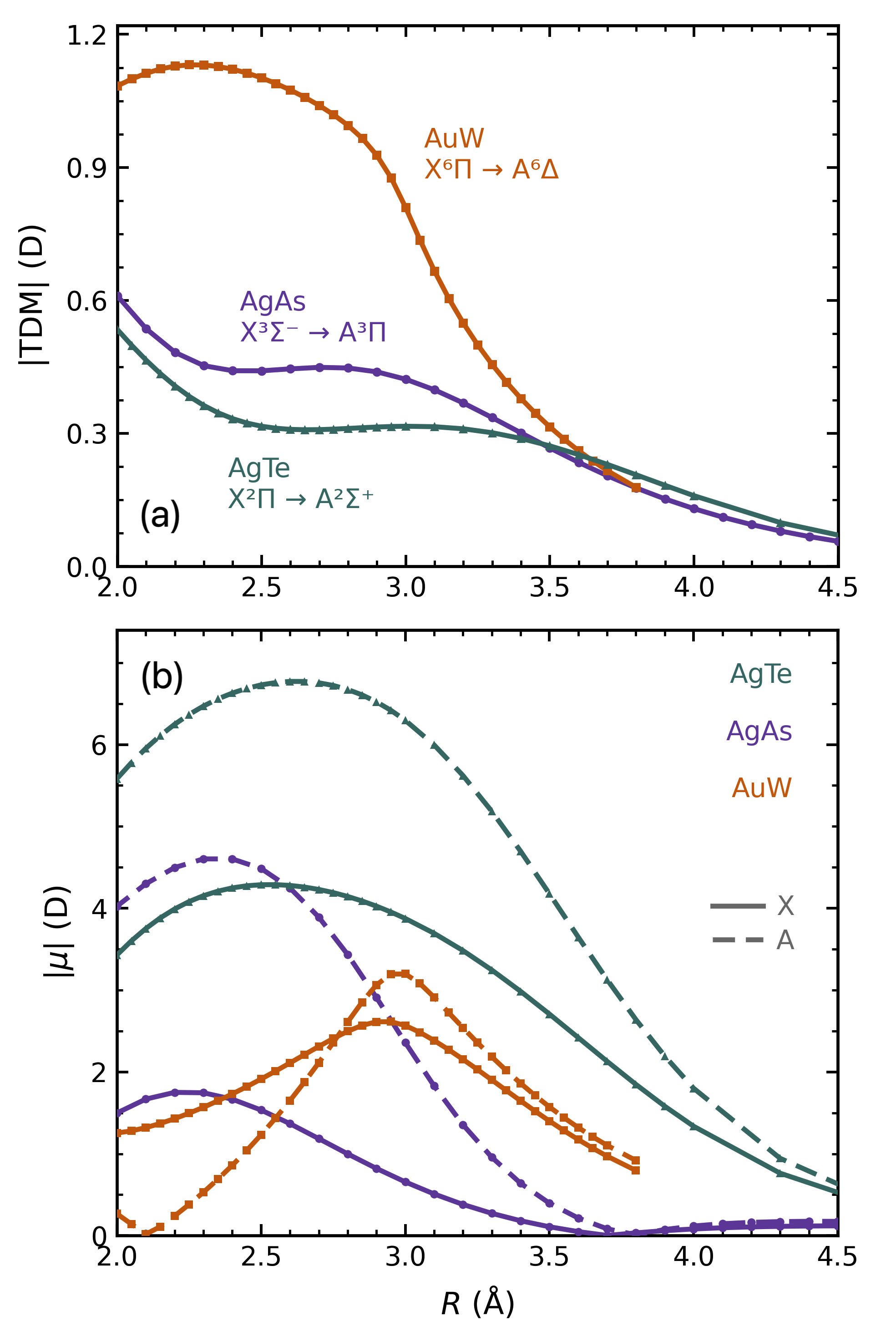}
  \caption{\label{fig:dipoles}
  Transition and permanent dipole moments of the coolable candidates and the MCSCF/MRCI+Q level. Panel (a) shows the transition dipole moment between the states of interest for the cooling transition of the three different candidates. Panel (b) displays the permanent dipole moments for the ground state (solid lines) and the cooling transition excited state (dashed lines)
}
\end{figure}

\subsection{Dipole moment calculations}
For completeness, we calculated the permanent dipole moments (PDMs) and transition dipole moments (TDMs) of the three candidates at the MRCI+Q level. Figure \ref{fig:dipoles} shows the TDMs and PDMs for the electronic states involved in the diagonal transitions. The calculated TDMs for AgAs and AgTe are of the same order of magnitude as those reported for AsH \cite{Li2022} and, in general, are smaller than those reported for the alkaline-earth-metal halides that have been successfully laser cooled to date \cite{Hao2019,Zhuang2011}. The TDMs of AuW are larger at short internuclear distances; however, the difference is relatively small, and their magnitudes remain below those reported for the alkaline-earth-metal halides. In contrast, the PDMs exhibit considerably larger values, particularly for AgTe, whose ground-state PDM reaches approximately 4 D near the equilibrium distance.

\begin{figure}[t]
  \includegraphics[width=\linewidth]{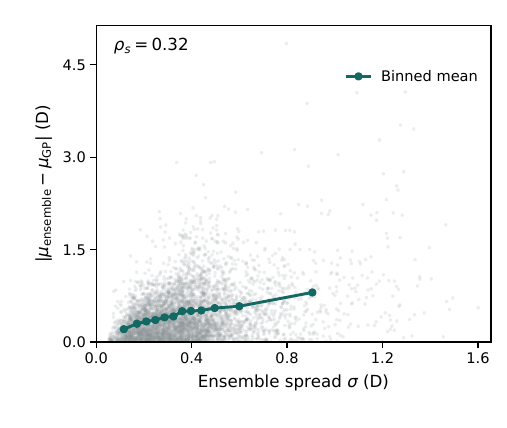}
  \caption{\label{fig:si-calib}
  Absolute difference between ensemble and Gaussian-process dipoles against
  ensemble spread for the $4{,}851$ molecules of the pretraining set. Grey points show
  individual molecules; the teal curve shows the mean in $12$ equal-population
  bins. The Spearman coefficient is $\rho_s=0.32$.}
\end{figure}

\begin{table}[h!]
\centering
\begin{threeparttable}
\caption{Spectroscopic Constants of AuW, AgAs and AgTe.  }
\setlength{\tabcolsep}{3.8pt}
\begin{tabular}{@{}lccccc@{}}
\toprule
\textbf{State} & $R_e$ (\AA) & $\omega_e$ (cm$^{-1}$) & $B_e$ (cm$^{-1}$)  & $T_e$ (cm$^{-1}$) \\
\midrule
\\
\multicolumn{6}{c}{\textbf{AuW}} \\
\\
X$^6\Pi$ & 2.6879 & 158.1& 0.0245 & 0 \\
\\
A$^6\Delta$      & 2.7267 & 154.2 & 0.0238 & 2597.0 \\
\midrule
\midrule
\multicolumn{6}{c}{\textbf{AgAs}} \\
\\
X$^3\Sigma^-$ & 2.4688 & 200.45 & 0.0628 & 0 \\
              & 2.416$^{\rm{a}}$  & 215$^{\rm{a}}$ &       &  \\
              \\
A$^3\Pi$      & 2.4636 & 186.02 & 0.0631&  15555.9 \\
\midrule
\midrule

\multicolumn{6}{c}{\textbf{AgTe}} \\
\\
X$^2\Pi$ & 2.574 & 191.60 & 0.0434  & 0 \\
              & 2.525$^{\rm{b}}$  &183.3$^{\rm{b}}$ &  0.0452$^{\rm{b}}$   & \\
              & &195$^{\rm{c}}$ &  & \\
              \\
A$^2\Sigma^+$ & 2.593 & 172.90 & 0.0427  & 13604.3 &  \\
 & 2.538$^{\rm{b}}$  &161.2$^{\rm{b}}$ &  0.0448$^{\rm{b}}$  & 11852$^{\rm{b}}$   \\
\bottomrule
\label{tab:spectroscopy}
\end{tabular}
\begin{tablenotes}
\footnotesize
\item[a] Ref. \cite{Carta2014}
\item[b] Ref. \cite{Abdulal2024}
\item[c] Ref. \cite{Maheshwari1963}
\end{tablenotes}

\end{threeparttable}
\end{table}

\subsection{Lifetimes and Franck--Condon factors calculations}
We use the potential-energy curves to compute the vibrational wavefunctions
and the transition dipole moments to estimate the radiative lifetimes.
The Franck--Condon factors are
\begin{equation}
 f_{\nu\nu'}
 =
 \left|
 \left\langle
 \psi_{\nu}^{\mathrm{g}}
 \middle|
 \psi_{\nu'}^{\mathrm{e}}
 \right\rangle
 \right|^2,
 \label{eq:fcf}
\end{equation}
where $\nu$ labels a vibrational level of the lower electronic
state and $\nu'$ labels a vibrational level of the upper
electronic state. The radiative lifetime is
\begin{equation}
 \tau_{\nu'}
 =
 \left(\sum_{\nu}A_{\nu,\nu'}\right)^{-1}.
 \label{eq:radiative-lifetime}
\end{equation}
Here $A_{\nu,\nu'}$ is the Einstein coefficient for spontaneous
emission from upper-state level $\nu'$ to lower-state level $\nu$. The vibrational states were computed using the Colbert--Miller DVR approach and were also implemented in the calculation of the Franck--Condon factors.

\begin{table}[t]
\caption{Radiative properties of the proposed cooling transitions. $\tilde{\nu}$ is the transition wavenumber, $\tau$ the radiative lifetime of the upper state,
$\Gamma/2\pi = 1/(2\pi\tau)$ the natural linewidth and $\mathcal{R}$ denotes the $\tau / \tau_{\rm{AsH}}$ ratio. Lifetimes are estimated using the transition dipole moments of Fig.~\ref{fig:dipoles} and the DVR vibrational states.}
\label{tab:lifetimes}
\begin{ruledtabular}
\begin{tabular}{lccccc}
Molecule & Transition & $\tilde{\nu}$ (cm$^{-1}$) & $\tau$ & $\Gamma/2\pi$ (kHz)
         & $\mathcal{R}$ \\
\hline
AsH  & $X\,^3\Sigma^- \rightarrow A\,^3\Pi$  & 30\,600 & 914~ns\footnotemark[1] & 174 & 1   \\
AgAs & $X\,^3\Sigma^- \rightarrow A\,^3\Pi$  & 15\,556 & 13.9~$\mu$s              & 11  & 15.2  \\
AuBi & $X\,^3\Sigma^- \rightarrow A\,^3\Pi$  & 17\,000 & 3.3~$\mu$s              &  48  &  3.6 \\
AgTe & $X\,^2\Pi \rightarrow A\,^2\Sigma^+$  & 13\,604 & 21~$\mu$s              & 7   & 23  \\
AuW  & $X\,^6\Pi \rightarrow A\,^6\Delta$    & 2\,597  & 0.58~ms                & 0.3 & 630 \\
\end{tabular}
\end{ruledtabular}
\footnotetext[1]{Ref. \cite{Li2022}}
\end{table}

Table \ref{tab:lifetimes} tabulated the obtained excited-state lifetimes, the associated natural linewidth, and the $\tau / \tau_{\rm{AsH}}$ ratio in order to compare the possible laser cooling efficiency of the candidates with respect to that of the previously addressed AsH \cite{Li2022}.

\section{Chemical calculations for the four clusters}
\label{app:chem-clusters}

A sample of five representative molecules in each cluster is
listed in Table~\ref{tab:clusters}. Their equilibrium bond lengths, vibrational
frequencies, dissociation energies, polarizabilities, bond orders, and dipole
moments are from the molecular dataset or calculated using the methods given
in Appendix~\ref{app:quantum-chemistry}.

\begin{table*}[t]
\caption{\label{tab:clusters}
Chemical properties of the four clusters, illustrated by a sample of five
molecules from each. The clusters are A metallic
(alkali), B covalent, C metallic (alkali-free), and D ionic; their construction is
given in Methods.
$q_{\mathrm{eff}}=|\mu|/(e_{\mathrm{el}}R_e)$ is the effective charge,
$\bar{\alpha}=(\alpha_{\parallel}+2\alpha_{\perp})/3$ is the mean
polarizability, and BO is the Mulliken bond order. For the Zn+AM(AEM)
molecules, the spectroscopic constants were taken from \cite{Klaudia2021},
the rest of the values presented come from the molecular dataset and
from the electronic-structure calculations described in
Appendix~\ref{app:quantum-chemistry}.}
\begin{ruledtabular}
\begin{tabular}{clrrrrrr}
Cluster & Molecules & $R_e$ (\AA) & $\omega_e$ (cm$^{-1}$) &
$D_e$ (eV) & $\bar{\alpha}$ (a.u.) & BO & $q_{\mathrm{eff}}$ \\
\colrule
A & NaRa, NaBa, NaSr, KBa, KSr
  & 3.9--4.5 & 54--94 & 0.15--0.27 & 398--620
  & 0.14--0.39 & 0.004--0.073 \\
B & SiO, NO, GeO, GeS, PN
  & 1.15--2.01 & 576--1904 & 5.7--8.3 & 11--54
  & 1.83--2.86 & 0.03--0.43 \\
C & ZnRb, ZnBa, ZnK, ZnSr, ZnRa
  & 4.0--4.5 & 22--35 & 0.046--0.055 & 250--374
  &0.02-0.15 & 0.001--0.009 \\
D & KCl, NaCl, KBr, RbCl, RbBr
  & 2.4--2.9 & 169--366 & 3.9--4.3 & 35--49
  & 0.11--0.23 & 0.768--0.799 \\
\end{tabular}
\end{ruledtabular}
\end{table*}

The calculated dipole moments and polarizabilities for clusters A and D are compared
with Refs.~\cite{Klaudia2021,Hela2024}.

% \textcolor{red}{TODO}

\paragraph*{Bonding clusters}
\label{app:cluster-method}

Table~\ref{tab:cluster-members} lists, for each cluster, the ten labeled molecules
of largest membership probability. The covalent cluster is the most sharply defined,
with $39$ molecules on which all $50$ members agree; the three others have no molecule
of unit probability except the sodium--alkali pairs of cluster~A, and their median
probabilities of $0.70$--$0.78$ reflect the softer boundaries between the metallic
halves and between the two polar regimes. The representatives of
Table~\ref{tab:clusters} in clusters A, B, and D have probabilities between $0.82$
and $1.00$.

\begin{table*}[t]
\caption{\label{tab:cluster-members}
Labeled molecules of largest membership probability in each cluster. The
probability $P$ is the fraction of the $50$ ensemble members that place the molecule
in its final cluster (Appendix~\ref{app:cluster-method}); molecules are ordered by
$P$ and, at equal $P$, by distance to the cluster center. The median $P$ is taken
over all molecules of the cluster. Pairs stored in both atom orders in the labeled
set receive the same assignment and are listed once.}
\begin{ruledtabular}
\begin{tabular}{clcl}
Cluster & & Median $P$ &
\parbox[t]{0.68\textwidth}{\raggedright Ten molecules of largest $P$} \\
\colrule
A & metallic (alkali) & 0.78 &
\parbox[t]{0.68\textwidth}{\raggedright
NaFr (1.00), NaK (1.00), NaCs (1.00), LiNa (0.96), LiBa (0.94), LiRb (0.94),
NaCa (0.94), NaBa (0.94), LiSr (0.94), NaSr (0.94)} \\
B & covalent & 0.98 &
\parbox[t]{0.68\textwidth}{\raggedright
FeO (1.00), PtC (1.00), GeO (1.00), NbO (1.00), PO (1.00), IrC (1.00),
CoO (1.00), SnO (1.00), SiO (1.00), TiS (1.00)} \\
C & metallic (alkali-free) & 0.70 &
\parbox[t]{0.68\textwidth}{\raggedright
CrSr (0.84), CrMg (0.84), CrCa (0.82), CrBa (0.82), CdCa (0.80), CdBa (0.80),
CdRa (0.80), CdMg (0.80), ZnRa (0.80), ZnMg (0.80)} \\
D & ionic & 0.76 &
\parbox[t]{0.68\textwidth}{\raggedright
KF (0.88), NaF (0.88), RbF (0.88), CsF (0.88), RbCl (0.86), LiH (0.86),
NaH (0.86), NaCl (0.84), NaBr (0.84), KCl (0.84)} \\
\end{tabular}
\end{ruledtabular}
\end{table*}

\section{Uncertainty and anomalous molecules}
\label{app:uncertainty}

The uncertainties are calibrated as described in Methods. For $D_e$,
$R_e$, and $\omega_e$, this gives $c_{68}=0.86$, $0.55$, and $0.64$,
respectively.

For the anomaly analysis, five quantities are calculated for each molecule: the
error of the predicted dipole and its uncertainty, the difference between learned and VAK gate
weights, the magnitude of the leading residual direction of the full representation
after a linear regression on $(\bar{\chi},\Delta\chi)$, and distance to the nearest
cluster center. Each quantity is converted to a percentile rank, and the five ranks are
averaged.

The highest-ranked anomalies are LiH, CO, CsF, CH, OH, NH, LaO, AlCl, MoC, and LiO. This ranking combines prediction residuals, uncertainty, and departures from the learned bonding trends; it does not identify a single shared chemical mechanism.

\begin{figure*}[t]
  \includegraphics[width=\textwidth]{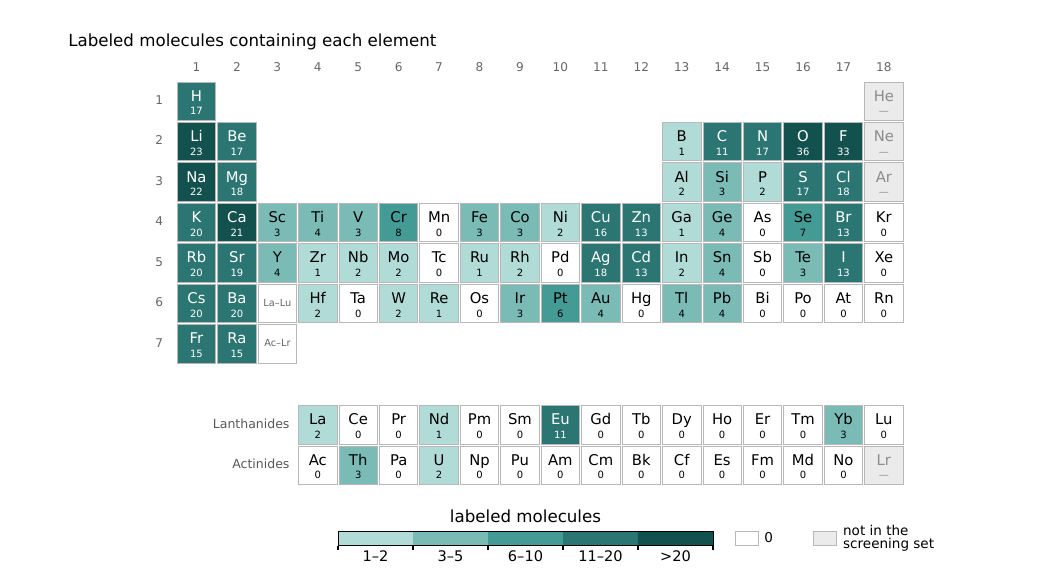}
  \caption{\label{fig:si-coverage}
  Number of the $290$ labeled molecules that contain each element. White tiles mark
  elements of the pretraining set that appear in no labeled molecule, and grey tiles mark
  elements outside the pretraining set. Of the $99$ elements in the pretraining set, $37$
  appear in no labeled molecule, among them Kr, Xe, and Rn, the heavy pnictogens As, Sb,
  and Bi, and most lanthanides and actinides. These elements enter $2{,}960$ of the
  $4{,}851$ molecules of the pretraining set.}
\end{figure*}

On the pretraining set, the comparison is between the ensemble and the
Gaussian-process values used during pretraining. It is therefore a comparison
between two models and not an independent test. The mean absolute difference is
$0.455$~D, with Pearson correlation $r=0.938$ and $R^2=0.878$. The largest differences occur
for alkali and alkaline-earth compounds of silicon, sulfur, selenium, and oxygen
(SrS, RbSi, SrSe, SrO, FrSi), combinations that are sparse in the labeled set.
Figure~\ref{fig:si-coverage} shows how the labeled molecules cover the periodic table.

\begin{figure*}[t]
  \includegraphics[width=\textwidth]{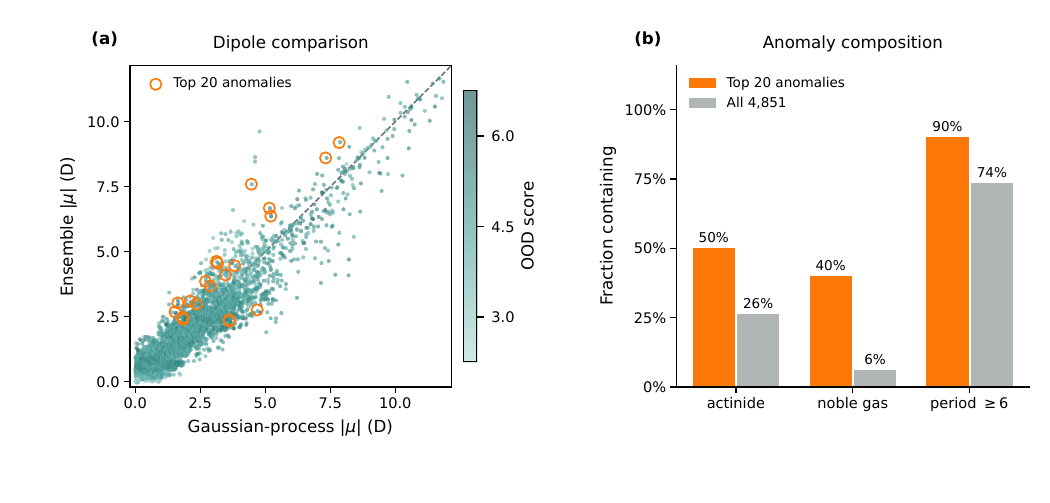}
  \caption{\label{fig:si-largeset}
  Comparison over the pretraining set. (a) Ensemble dipole
  against the Gaussian-process value used during pretraining, colored by the
  out-of-distribution score. Circled points are the largest anomalies. This is a
  comparison between two models and is not an external test. (b) The largest
  anomalies are mainly heavy molecules containing actinides or noble gases, and
  hydrides.}
\end{figure*}

\section{Ions and charge transfer}
\label{app:ions}

The extension to charged species uses a separate model that includes the molecular
charge as an input. It is not the neutral model discussed in the main text. The
charged model contains $12$ atomic and molecular inputs and is trained jointly on
neutral molecules and ions. Missing ion dipoles are masked in the loss.

A five-fold cross-validation over $81$ ions gives $R^2=0.95$ for $R_e$ and
$R^2=0.94$ for $\omega_e$. The corresponding values for $D_e$ and the dipole are
$0.55$ and $0.64$. Over the $66$ cations, the values are $0.95$, $0.95$, $0.53$,
and $0.64$ for $R_e$, $\omega_e$, $D_e$, and the dipole, respectively.

The in-sample values over all $81$ ions are $0.99$, $0.99$, $0.94$, and $0.93$.
The difference between the in-sample and cross-validated values shows that the
energetics and dipoles are partly memorized. The ion dataset contains only
$15$ anions, which limits comparisons across charges.

\begin{figure*}[t]
  \includegraphics[width=\textwidth]{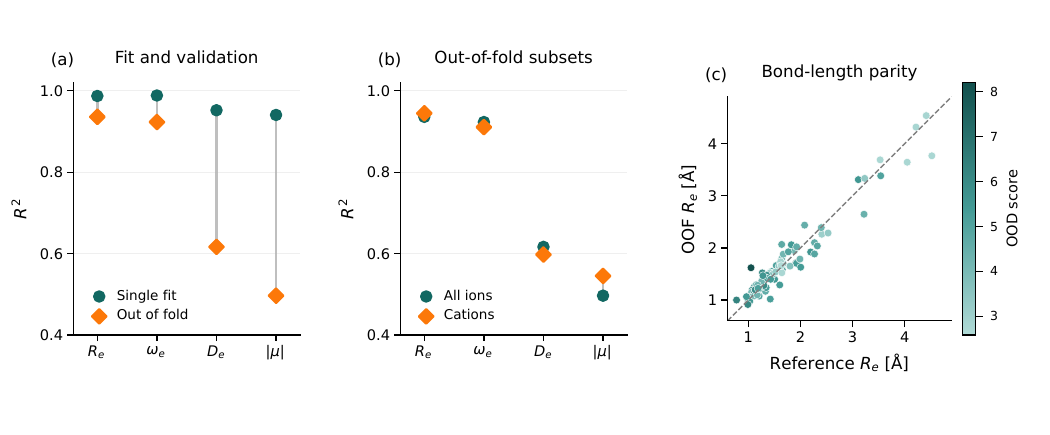}
  \caption{\label{fig:si-ions}
  Extension to ions. (a) In-sample and out-of-fold $R^2$ for each molecular
  quantity. Bond length and vibrational frequency transfer more accurately than
  dissociation energy and dipole. (b) Out-of-fold comparison with and without the
  anions. (c) Bond-length parity for the $81$ ions, colored by the
  out-of-distribution score.}
\end{figure*}

\section{Interpretability of the residual coordinate}
\label{app:symbolic}

The residual coordinate $\xi$ contains the part of the symmetric representation that
is not described by the VAK coordinates [Eqs.~\eqref{eq:vak-residual} and
\eqref{eq:xi}]. Its meaning must therefore come from the atoms that form each molecule.
We examine it in three steps. We first compare $\xi$ with the atomic quantities, then
separate the contribution of each element, and finally look for closed-form
expressions that reproduce it.

\begin{figure*}[t]
  \includegraphics[width=\textwidth]{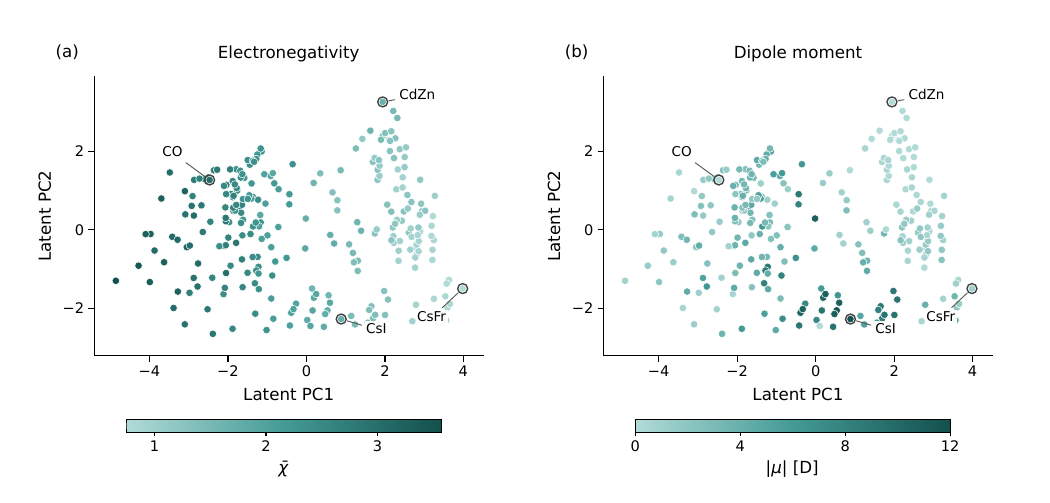}
  \caption{\label{fig:si-latent-pca}
  Leading principal-component plane of the symmetric representation for
  the $262$ molecules of the training set, colored by (a) mean electronegativity
  $\bar{\chi}$ and (b) reference dipole $|\mu|$. Both panels use identical
  coordinates and scales.}
\end{figure*}

Figure~\ref{fig:si-latent-pca} shows the leading linear projection of the symmetric
representation.

\begin{figure*}[t]
 \includegraphics[width=\textwidth]{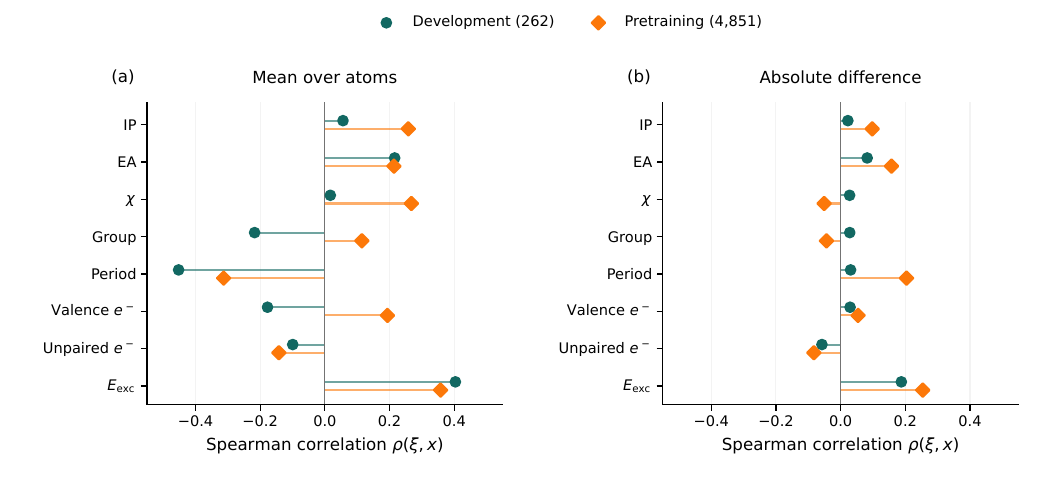}
 \caption{\label{fig:xi-correlations}
 Rank correlation of the residual coordinate $\xi$ with the atomic quantities, for the
 labeled molecules and for the $4{,}851$ molecules of the pretraining set. (a) Mean
 over the two atoms. (b) Absolute difference between them.}
\end{figure*}

Figure~\ref{fig:xi-correlations} shows the rank correlation of $\xi$ with the atomic
quantities. For both the labeled molecules of Fig.~\ref{fig:plane} and the $4{,}851$
molecules of the pretraining set, $\xi$ decreases with the period of the atoms and
increases with their first excitation energy. The electron affinity gives a weaker
positive trend. The group, the valence-electron count, and the ionization potential
behave differently in the two sets, and the differences between the two atoms
correlate less with $\xi$ than their mean. The electronegativity is uncorrelated with
$\xi$ for the labeled molecules, on which the VAK regression is fitted. Over the
pretraining set, its correlation with $\bar{\chi}$ is $0.27$.
The contribution of each element follows from the construction of $\xi$. The
symmetric channel is the mean of the two atomic encodings,
$\mathbf{z}_{\mathrm{sym}}=\tfrac{1}{2}(\mathbf{h}_A+\mathbf{h}_B)$, and $\xi$ is a linear projection of
$\mathbf{z}_{\mathrm{sym}}-\widehat{\mathbf{z}}_{\mathrm{VAK}}(\bar{\chi},\Delta\chi)$. The coordinate
therefore separates exactly into a term for each atom and a function of the VAK
coordinates,
\begin{equation}
 \xi(AB)=\tfrac{1}{2}\big[e(A)+e(B)\big]+g(\bar{\chi},\Delta\chi).
 \label{eq:xi-template}
\end{equation}
Here $e(X)$ is the projection of the encoding $\mathbf{h}_X$ of atom $X$, and $g$ is minus the
projection of the VAK regression, up to a constant. The value of $e$ for every element
can be read directly from the model. It is large for lithium, sodium, the coinage
metals, and the halogens. These atoms have a single electron
outside a closed shell or a single vacancy in one. Among the alkali metals, $e$
decreases with the period, and caesium and francium lie near the middle of the
distribution. The smallest values occur for zinc and cadmium, with filled $d^{10}s^2$
valence shells, and for the lanthanides, with
their open $f$ shells. The noble gases, the $p$-block elements, and most transition
metals lie in between. The halogens and the alkali metals are at opposite ends of the
electronegativity scale, yet both raise $\xi$. This ordering is therefore not
contained in the electronegativity. Moreover, this ordering holds across all
pretrained models.

We then search for closed-form expressions for $e$ and $g$ using symbolic regression.
Symbolic regression represents candidate expressions as trees of operations on a
chosen set of variables, and evolves them to balance accuracy against length. We use
PySR~\cite{pysr_paper}, whose expression templates allow the form of
Eq.~\eqref{eq:xi-template} to be imposed. The same function $e$ is applied to both
atoms, and $g$ depends only on $\bar{\chi}$ and $\Delta\chi$. The function $e$ uses the
ionization potential IP, electron affinity EA, first atomic excitation energy
$E_{\mathrm{exc}}$, group $G$, period $P$, and numbers of valence and unpaired
electrons, $V$ and $U$, with energies in eV. The encoding of an atom also depends on
its electronegativity, but we leave $\chi$ out of $e$ so that the whole
electronegativity dependence is carried by $g$. The available operations are addition,
subtraction, multiplication, the square, the exponential, and the hyperbolic tangent.
We exclude division to avoid singular expressions, and $e$ and $g$ together contain at
most $60$ symbols. The expressions are fitted to the values of $\xi$ for the
pretraining set, divided by atom pair into training, validation, and test molecules in
the proportion $3{:}1{:}1$.

For each length, a search returns the most accurate expression. We select among these
with the default criterion of PySR. Of the expressions whose loss is within a factor
$1.5$ of the most accurate one, it takes the expression with the largest decrease in
the logarithm of the loss per additional symbol. Three independent searches reproduce
$\xi$ for the test molecules with $R^2$ between $0.70$ and $0.78$, compared with
$0.96$--$0.97$ for flexible regressions on the same atomic quantities. The atomic
quantities are strongly correlated across the periodic table, so the searches reach
this accuracy through equivalent combinations of them. All of these combinations
contain the valence-electron count. One of the selected expressions is
\begin{align}
 e &= \tanh\!\Big\{(2.15-0.354\,G)^4\Big[\big(E_{\mathrm{exc}}\,\mathrm{EA}^2\,\\&(2.82-0.181\,G)\big)^2-1.81\Big]\Big\}
   +13.0\,(1-\tanh V),\nonumber\\
 g &= 1.44\,\bar{\chi}+\tanh\!\big[\Delta\chi-(1.81\,\Delta\chi+\bar{\chi}-1.87)^2\big]-2.83,
 \label{eq:xi-sr}
\end{align}
which reproduces $\xi$ for the test molecules with $R^2=0.76$. Its last term raises the
atoms with a single valence electron. Its first term separates the atoms by a
group-dependent threshold on the product of the excitation energy and the squared
electron affinity.

In agreement with the atomic contribution, this expression implies that the coordinate $\xi$ follows the valence occupancy of the atoms. It
is raised by a single electron outside a closed shell or a single vacancy in one, and
lowered by filled
$s$ and $d$ valence subshells and by open $f$ shells. This information is not
described by the electronegativity.

\section{Predicted properties of the laser-cooling candidates}
\label{app:candidate-properties}

Table~\ref{tab:candidate-properties} presents the predicted ground-state properties
and cluster-membership probabilities for the candidate molecules of
Table~\ref{tab:shortlistFCF}, together with the other
candidates (CuSb, CuBe, PaO, SnF, TcF). All entries are means over the
$50$-member ensemble. The dipole moment is the decoder output; $D_e$, $R_e$, and
$\omega_e$ are the Gaussian-process predictions from the representation
(Appendix~\ref{app:validation}). The cluster probabilities
$P_{\mathrm{A}}$--$P_{\mathrm{D}}$ are obtained by $k$-nearest-neighbor voting
($k=5$) across the ensemble so that within each member's latent space a $k$-NN classifier
trained on the labeled partition votes for a cluster, and the votes are averaged over
the members, the same construction used for the cluster representatives.

Each property is quoted with its calibrated $68\%$ half-width
(Appendix~\ref{app:uncertainty}). The bond length is determined to a few percent for every
candidate ($0.05$--$0.13$~\AA). The other constants separate the two chemical
families in the table. For the fluorides and oxides the interval on $\omega_e$ is
$10$--$25\%$ of the value ($1012(100)$~cm$^{-1}$ for PaO, $632(77)$ for SnF,
$882(203)$ for TcF). For the coinage-metal--pnictogen and --chalcogen pairs it is
$35$--$45\%$ ($297(103)$~cm$^{-1}$ for AgAs, $246(109)$ for AuBi, $266(108)$ for
CuSb), and the dipole and $D_e$ carry intervals of $20$--$45\%$. Of the $262$ training molecules, $63$ contain fluorine or oxygen, whereas none
pairs a coinage metal with a pnictogen, or with a chalcogen heavier than sulfur. The
members are
therefore interpolating in the first region and extrapolating in the second, and the
reported intervals measure how well the training data cover each species.

\begin{table*}[t]
\caption{\label{tab:candidate-properties}
Ensemble predictions for the laser-cooling candidates of Table~\ref{tab:shortlistFCF}
and the other candidates:
dipole moment $|\mu|$ (D), dissociation energy $D_e$ (eV), equilibrium bond length
$R_e$ (\AA), and vibrational frequency $\omega_e$ (cm$^{-1}$), each a posterior mean
over the $50$-member ensemble. The last four columns are the cluster-membership
probabilities $P_{\mathrm{A}}$--$P_{\mathrm{D}}$ (the $k$-nearest-neighbor vote of
Appendix~\ref{app:cluster-method}), which sum to one across the row; the largest entry
is the model's assigned cluster, given in column ``Cl.'' A value near one (AgTe, PaO,
SnF) is a confident assignment, whereas probability spread across clusters (AgAs) marks
a weaker one. The dipole is the decoder prediction and $D_e$, $R_e$, $\omega_e$ are the
Gaussian-process predictions from the representation (Appendix~\ref{app:validation}).
Parenthesised digits are the calibrated $68\%$ half-width in units of the last quoted
digit: $3.61(116)$~eV means $3.61\pm1.16$~eV at $68\%$ confidence
(Appendix~\ref{app:uncertainty}).}
\begin{ruledtabular}
\begin{tabular}{llcccrcccc}
Molecule & Cl. & $|\mu|$~(D) & $D_e$~(eV) & $R_e$~(\AA) & $\omega_e$~(cm$^{-1}$) & $P_{\mathrm{A}}$ & $P_{\mathrm{B}}$ & $P_{\mathrm{C}}$ & $P_{\mathrm{D}}$ \\
\colrule
AsH  & D & 1.41(34) & 3.61(116) & 1.42(7) & 2291(196) & 0.00 & 0.00 & 0.00 & 1.00 \\
AuW  & B & 3.53(138) & 2.71(103) & 2.49(13) &  270(110) & 0.07 & 0.92 & 0.00 & 0.02 \\
AgAs & B & 2.48(83) & 3.39(68) & 2.32(7) &  297(103) & 0.20 & 0.78 & 0.00 & 0.01 \\
AgTe & B & 2.77(74) & 2.31(67) & 2.49(5) &  237(82)  & 0.03 & 0.96 & 0.00 & 0.01 \\
AuBi & B & 3.02(84) & 3.43(69) & 2.53(11) &  246(109) & 0.04 & 0.96 & 0.00 & 0.00 \\
CuSb & B & 1.63(71) & 3.14(74) & 2.40(8) &  266(108) & 0.11 & 0.88 & 0.00 & 0.00 \\
CuBe & A & 1.00(72) & 2.12(111) & 2.44(12) & 413(147) & 0.98 & 0.00 & 0.02 & 0.00 \\
PaO  & B & 2.69(42) & 8.48(95) & 1.69(6) & 1012(100) & 0.00 & 1.00 & 0.00 & 0.00 \\
SnF  & D & 3.20(71) & 5.15(60) & 1.88(5) &  632(77)  & 0.00 & 0.00 & 0.00 & 1.00 \\
TcF  & D & 2.26(108) & 5.71(94) & 1.71(9) &  882(203) & 0.00 & 0.00 & 0.00 & 1.00 \\
\end{tabular}
\end{ruledtabular}
\end{table*}

\section{Most anomalous molecules}
\label{app:anomalous}

The model flags unusual molecules through two independent views. Table~\ref{tab:anom-latent}
lists the largest latent-space anomalies over the full pretraining set,
ranked by the distance of the ensemble-mean latent position to the nearest cluster
centroid (the latent out-of-distribution score). Table~\ref{tab:anom-error} lists
the largest out-of-fold dipole errors within each cluster over the labeled set.

The member-to-member
spread $\sigma_{\mu}$ measures how much the answer depends on which subset of the data a
member saw; calibrated as in Appendix~\ref{app:uncertainty}, it gives the $68\%$
intervals quoted in the tables.

For the latent anomalies, the spread $\sigma_{d}$ of the nearest-centroid distance
across members is small compared with the distances themselves
($\sigma_{d}\!\approx\!0.4$--$0.8$ against $d_{\mathrm{latent}}\!\approx\!6.3$--$6.8$,
versus a full-set median $d_{\mathrm{latent}}=4.25$). The members therefore agree
that these molecules are outliers, so the anomaly is a property of the learned
representation, shared by all members. Their dipole intervals,
$0.56$--$1.27$~D at $68\%$, are comparable to or somewhat wider than the full-set
median of $0.56$~D, so the flag marks a region of larger uncertainty.

For the residual anomalies the informative quantity is the residual in units of the
calibrated interval, $r_{\mathrm{norm}}=|\Delta\mu|/(c_{68}\sigma_{\mu})$. Over the labeled set the
median is $r_{\mathrm{norm}}=0.69$, as expected for a $68\%$ interval, and the largest errors in each
cluster reach $r_{\mathrm{norm}}=1.0$--$2.2$, that is, one to two intervals. The largest errors are the chromium compounds
CrBa and CrMg in cluster~C, the metal--metal pairs AgCu and LiBe in cluster~A, and
the heavy oxide, sulfide, and carbide SrO, BaS, and MoC in cluster~B, combinations
for which the labeled set holds few neighbors.

\begin{table}[t]
\caption{\label{tab:anom-latent}
Largest latent-space anomalies over the pretraining set: distance of the
ensemble-mean latent to the nearest cluster centroid ($d_{\mathrm{latent}}$), with
the predicted cluster and dipole. The extreme is dominated by hydrides and by
caesium and heavy-actinide compounds, combinations sparsely represented in the
labeled set. $\sigma_{d}$ is the standard deviation of the
same distance recomputed within each of the $50$ ensemble members separately (each
member using its own cluster centroids), and the last column the calibrated $68\%$
half-width of the dipole. Both are small relative to the anomaly itself, so the members
agree on the flag. Full-set
medians for reference: $d_{\mathrm{latent}}=4.25$, $\sigma_{d}=0.46$.}
\begin{ruledtabular}
\begin{tabular}{lccccc}
Molecule & Cluster & $d_{\mathrm{latent}}$ & $\sigma_{d}$ & $|\mu|$~(D) & $68\%$~(D) \\
\colrule
CsH  & D & 6.76 & 0.68 & 6.60 & 1.27 \\
HPa  & D & 6.48 & 0.45 & 2.41 & 0.61 \\
CdH  & D & 6.47 & 0.72 & 1.67 & 0.96 \\
CfH  & D & 6.42 & 0.69 & 2.48 & 0.80 \\
BCs  & A & 6.39 & 0.65 & 3.06 & 1.17 \\
HZn  & D & 6.38 & 0.58 & 1.27 & 0.76 \\
HNo  & D & 6.32 & 0.68 & 2.40 & 0.85 \\
MoZn & B & 6.29 & 0.77 & 0.95 & 0.77 \\
AmH  & D & 6.28 & 0.60 & 2.49 & 0.63 \\
CsPa & A & 6.27 & 0.44 & 2.47 & 0.56 \\
\end{tabular}
\end{ruledtabular}
\end{table}

\begin{table}[t]
\caption{\label{tab:anom-error}
Largest out-of-fold dipole errors within each cluster over the labeled set:
reference $|\mu|_{\mathrm{ref}}$, ensemble out-of-fold prediction
$|\mu|_{\mathrm{pred}}$, and absolute error $|\Delta\mu|$, in Debye. The prediction
is quoted with its calibrated $68\%$ half-width, and the final column gives the
residual in units of that interval, $r_{\mathrm{norm}}=|\Delta\mu|/(c_{68}\sigma_{\mu})$. A value
above one means the reference lies outside the $68\%$ band; the largest errors reach
about twice it. The median over the labeled set is $r_{\mathrm{norm}}=0.69$.}
\begin{ruledtabular}
\begin{tabular}{llcr@{\,$\pm$\,}lcc}
Cluster & Molecule & $|\mu|_{\mathrm{ref}}$ & \multicolumn{2}{c}{$|\mu|_{\mathrm{pred}}$} & $|\Delta\mu|$ & $r_{\mathrm{norm}}$ \\
\colrule
A & AgCu &  0.04 &  1.86 & 1.17 & 1.83 & 1.6 \\
A & LiBe &  3.60 &  1.99 & 0.85 & 1.61 & 1.9 \\
A & AgBe &  0.71 &  1.80 & 0.85 & 1.09 & 1.3 \\
\colrule
B & SrO  &  8.90 &  6.63 & 1.43 & 2.27 & 1.6 \\
B & MoC  &  6.07 &  3.85 & 1.37 & 2.22 & 1.6 \\
B & BaS  & 10.86 &  9.09 & 1.37 & 1.77 & 1.3 \\
\colrule
C & CrBa &  2.67 &  1.51 & 0.68 & 1.16 & 1.7 \\
C & CrMg &  0.10 &  0.98 & 0.40 & 0.88 & 2.2 \\
C & CrBe &  1.43 &  1.03 & 0.41 & 0.40 & 1.0 \\
\colrule
D & YbF  &  3.91 &  2.20 & 1.03 & 1.71 & 1.7 \\
D & PtF  &  3.42 &  2.72 & 0.54 & 0.70 & 1.3 \\
D & NaH  &  6.40 &  5.72 & 0.68 & 0.68 & 1.0 \\
\end{tabular}
\end{ruledtabular}
\end{table}

\end{document}